\pdfoutput=1
\RequirePackage{ifpdf}

\documentclass{JINST}
\usepackage{cite}
\usepackage{url}

\title{Testing of High Voltage Surge Protection Devices for Use in Liquid Argon TPC Detectors}

\author{
J. Asaadi$^a$,
J.M. Conrad$^b$,
S. Gollapinni$^c$,
B.J.P. Jones$^b$\thanks{Corresponding Author},
H. Jostlein$^d$,
J.M. St. John$^e$,
T. Strauss$^f$,
S. Wolbers$^d$, and
J. Zennamo$^g$.\\
\llap{$^a$}Syracuse University,
   900 South Crouse Ave, Syracuse, NY 13244, USA \\ 
\llap{$^b$}Massachusetts Institute of Technology,
  77 Massachusetts Avenue, Cambridge, MA 02139, USA\\
  E-mail: \email{bjpjones@mit.edu}\\
\llap{$^c$}Kansas State University,
  Manhattan, KS 66506, USA\\
\llap{$^d$}Fermi National Accelerator Laboratory,
  Batavia, IL 60510, USA \\
\llap{$^e$}University of Cincinnati,
  2600 Clifton Ave, Cincinnati, OH 45220, USA  \\
\llap{$^f$}University of Bern,
  Albert Einstein Center, LHEP, Sidlerstasse 5, CH-3012 Bern, Switzerland \\
\llap{$^g$}University of Chicago, Enrico Fermi Institute, 
  5801 S Ellis Ave, Chicago, IL 60637, USA }

\abstract{In this paper we demonstrate the capability of high voltage varistors and gas discharge tube arrestors for use as surge protection devices in liquid argon time projection chamber detectors.  The insulating and clamping behavior of each type of device is characterized in air (room temperature), and liquid argon (90~K), and their robustness under high voltage and high energy surges in cryogenic conditions is verified.  The protection of vulnerable components in liquid argon during a 150 kV high voltage discharge is also demonstrated.  Each device is tested for argon contamination and light emission effects, and both are constrained to levels where no significant impact upon liquid argon time projection chamber functionality is expected.  Both devices investigated are shown to be suitable for HV surge protection applications in cryogenic detectors.}


\begin{document}

\tableofcontents

\section{Introduction} \label{sec:Introduction}

Large liquid argon time projection chambers (LArTPCs) are of great importance to the field of experimental neutrino physics~\cite{Raselli:2013oxa,LBNE,Chen:2007ae,deGouvea:2013onf,Laguna2013kaa}. The increasing scale of such detectors is motivated by the need to collect a large sample of neutrino interactions, often at a distance from a pulsed neutrino beam.  Liquid argon TPCs for direct dark matter detection~\cite{Zani:2014lea,Rielage:2012zz,Boulay:2012hq,Bossa:2014cfa,Badertscher:2013ygt} are also increasing in scale, a large fiducial volume being required to explore the parameter space of small dark matter interaction cross sections which the next generation of such experiments intends to probe.

\begin{figure}[t]
\centering \includegraphics[width=0.98\textwidth]{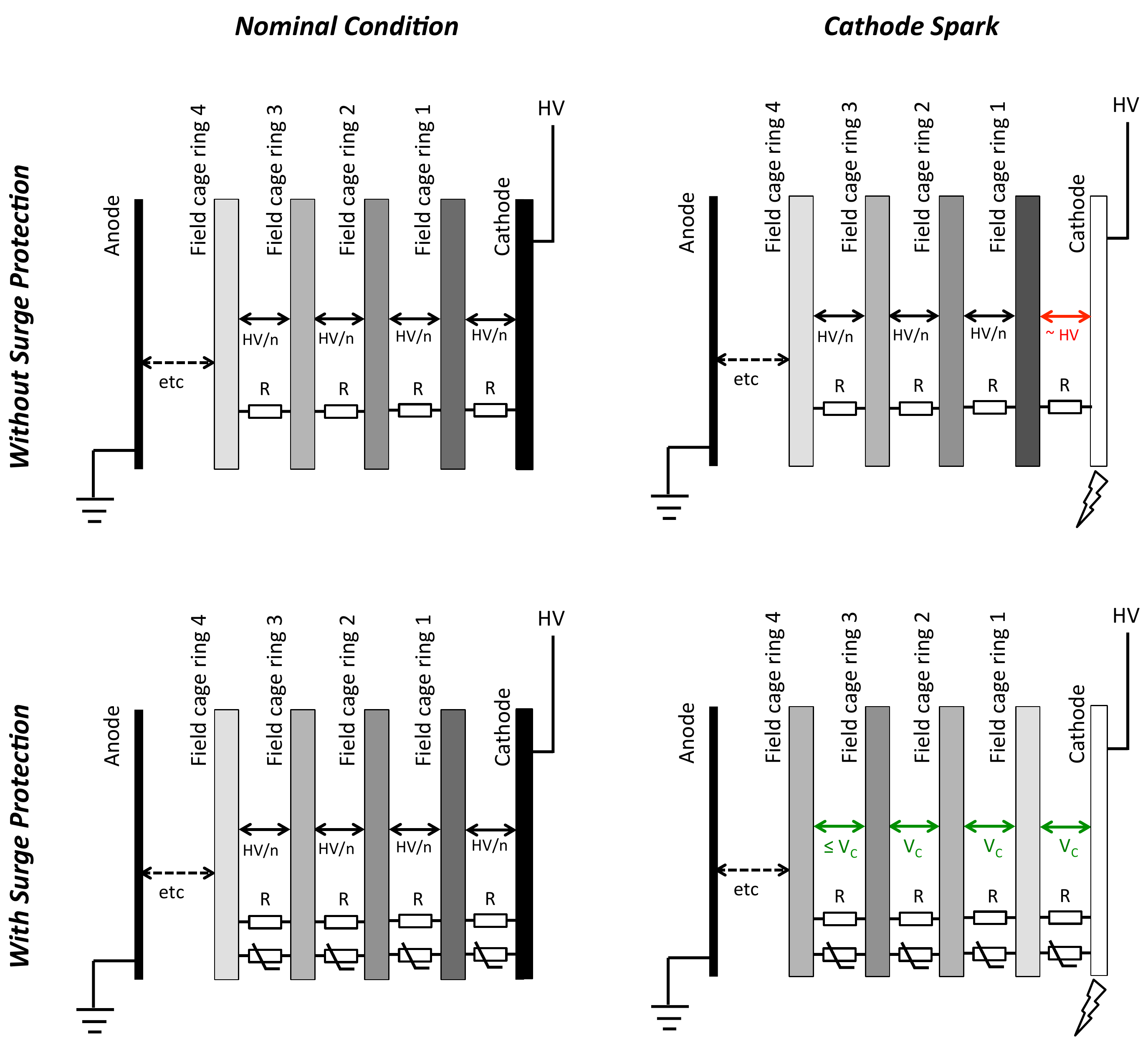}
\caption{A cartoon showing the behavior of the field cage potential divider in nominal (left) and fault (right) conditions, without (top) and with (bottom) surge protection.  Both with and without surge protection, in nominal running the applied high voltage is divided uniformly between the n field cage rings.  In the case of a spark from the cathode, a large transient over-voltage may exist for elements near the spark location, which can take a long time to disperse.  The scale of the over-voltage is determined by the field cage capacitances, with the full experiment HV being the worst case (shown).  With surge protection applied, the largest transient which can evolve is limited by the clamping voltage of the device.  The exact voltages evolved in each case depend on the details of the capacitances and geometry of the field cage. \label{fig:FaultConditionPic} } 
\end{figure}

To provide a strong enough electric field for electron drift over a large distance to readout wires near ground, a high voltage must be supplied to the cathode of the TPC.  To increase the drift length of a liquid argon TPC whilst maintaining a fixed electric field strength, the high voltage supplied must increase proportionally.  For the present generation of large LArTPCs, voltages of order 100~kV are required in order to provide a constant 500~V/cm electric field for approximately two meters.  To ensure uniformity, this field is usually shaped by a series of field cage rings which are connected by resistors, thus creating a voltage divider. This network is shown schematically in Figure~\ref{fig:FaultConditionPic}, top left.

The need for larger LArTPC detectors with higher applied voltages has led to new studies of the dielectric strength of liquid argon~\cite{HVC,Bern,Bay:2014jwa}.  In contrast to the commonly used reference value of 1.4~MV/cm from Ref.~\cite{Swan}, which was measured using only moderately pure argon, recent investigations have shown that the dielectric strength of argon at LArTPC purity with sub-ppb concentrations of oxygen and water may be much lower than anticipated, with Ref.~\cite{Bern} reporting breakdowns at 40~kV/cm.  In addition to argon purity, the breakdown voltage of liquid argon also depends upon electrode geometry, surface finish and accumulated space charge in ways which are not well understood at the present time.  In addition to implementing a LArTPC detector design which minimizes the likelihood of internal HV discharges occurring, it is also prudent to take steps to protect components which are vulnerable to damage in the event of an unexpected high voltage discharge.

High voltage breakdowns across the liquid argon volume can cause serious damage to electrical components in a TPC detector.  Components which may be at risk include not only those which are directly in the spark path, but also those which experience large potential differences some time after the spark has dissipated, due to transient over-voltages in the field cage.  For a large liquid argon TPC like MicroBooNE or LBNE, typical capacitances between elements of the field cage will be of order $0.1-1$~nF, whereas the resistances between some elements in the field cage network are of order $0.1-1$~G$\Omega$. This suggests that the relaxation time of the system after a high-voltage discharge may be of order $0.1-1$~s.  During this time, a significant fraction of the full high voltage of the experiment can be held between elements which would ordinarily only experience a moderate potential difference.  An over-voltage in case of a discharge arises due to the distribution of interelectrode capacitances given by the geometry of the electrode (cathode-field cage-cryostat) system.  The field cage voltage divider without and with surge protection is illustrated schematically in Figure~\ref{fig:FaultConditionPic}.

One method of protecting electrically sensitive components from large transient over-voltages is the application of a surge arrestor.  A surge arrestor is a device which is insulating at low voltages, but allows a current to flow if a critical voltage is exceeded.  This prevents large potential differences from being developed across a vulnerable component by allowing charge to be redistributed within the system more quickly than is allowed by the intrinsic resistances.  Commercially available surge protectors include varistors~\cite{pano} and gas discharge tubes (GDTs)~\cite{epcos}, both of which are used widely in various industrial applications. The commercially
available surge arrestors at the time of writing have not been rated for use at
liquid argon temperatures and none have been previously used in a running LArTPC detector.

In this paper we report on tests of high voltage GDTs and varistors for the protection of vulnerable components in the event of HV discharges in LArTPCs.  The specific application which initiated this investigation is the protection of field cage resistors for the MicroBooNE detector.  The requirements for this application are given below.  The derivation of these requirements based on the MicroBooNE TPC geometry and electrical properties, as well as the experimental testing performed to verify these properties, are outside the scope of this paper and will be reported in a future publication~\cite{uBooNE_NIM}.  Our goal is to demonstrate that both GDTs and varistors continue to function well in a cryogenic environment and do not have detrimental effects on key aspects of LArTPC operation, and as such are appropriate for both this specific application, and more general use cases in cryogenic TPC detectors.  The requirements for a surge protection device suitable for the MicroBooNE detector are:
\begin{itemize}
\item The device must have a significantly higher resistance than the field cage resistances of 250~M$\Omega$ at the nominal operating voltage of 2~kV.
\item The device must clamp the over-voltage that evolves during a fault condition to less than the voltage which can damage the field cage resistors in liquid argon, determined experimentally to be 30~kV.
\item The device must be able to survive repeated discharges of the system, which are estimated to deposit at most 2~J of energy per device
\item The device must function in cryogenic temperatures and in the dielectric environment of LArTPC purity liquid argon.
\item The device must not damage the argon purity in the detector
\item The device must not emit a sufficient light to interfere with the optical systems of the experiment in the nominal running condition
\end{itemize}

Section~\ref{sec:Devices} gives a brief discussion of the devices under test.  In Section~\ref{sec:BehaviorInCryogenics}, we demonstrate the functionality of GDTs and varistors in cryogenic environments and investigate how the insulating and clamping properties are altered in a liquid argon environment.  In Section~\ref{sec:RobustnessUnderSurges} we show how devices degrade under repeated surges when immersed in liquid argon and demonstrate their robustness in the expected surge conditions in a large LArTPC.  In Section~\ref{sec:PracticalConsiderations} we investigate the effects of the devices on argon purity and the ambient photon background.  Finally in Section~\ref{sec:Conclusions} we present our conclusions and briefly describe the surge protection solution implemented in the MicroBooNE detector.

\section{Devices under test} \label{sec:Devices}

\begin{figure}[t]
\centering \includegraphics[width=0.5\textwidth]{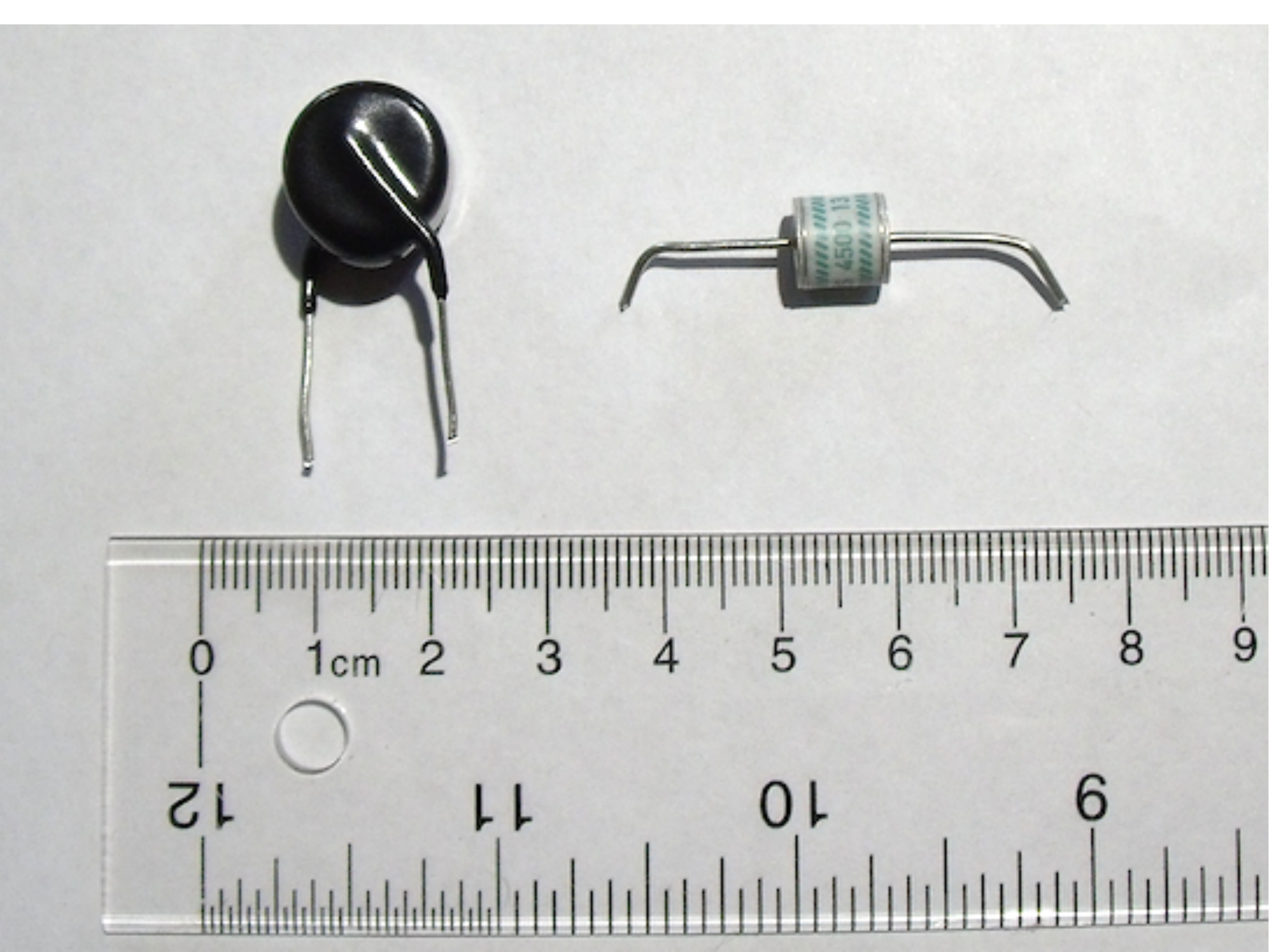}
\caption{Photograph of the devices under test.  Left: Panasonic ERZ-V14D182 varistor.  Right: EPCOS A71-H45X GDT \label{fig:IVCurves} } 

\end{figure}

\subsection*{Varistors} \label{sec:Varistors}

Varistors, also known as ``metal oxide varistors" (MOVs) and ``zinc oxide nonlinear resistors" (ZNRs), are commercially available surge protection devices, which use a pressed zinc oxide ceramic held between two electrodes.  Often the device is also encapsulated in a coating of epoxy, plastic or glass for electrical insulation.  The matrix of zinc oxide grains behaves as a large network of p$-$n semiconductor junctions.  This network produces a highly nonlinear I$-$V characteristic, with larger voltages giving lower effective resistances.  An example I$-$V curve is shown in Figure~\ref{fig:IVCurves}, left.  The varistor offers protection to sensitive components when it is applied in parallel with them, by allowing a large current to flow when there is an over-voltage.  This prevents large potential differences from developing over the vulnerable component, with the maximum voltage which can be developed being "clamped" at a lower value than in the unprotected circuit.  The voltage at which clamping occurs depends upon the available current, but the highly nonlinear I$-$V characteristic means the clamping voltage varies little over many orders of magnitude of applied currents.  In the literature and data sheets, the clamping or varistor voltage is reported as the applied voltage for which a specified current can flow, though the exact reference current varies from source to source.   For the purposes of this paper, we will define the clamping voltage as the applied voltage when 0.5~mA of current can flow.  For more information on the physics and functionality of varistors, see Ref.~\cite{var_material_science}.

\begin{figure}[t]
\centering \includegraphics[width=0.46\textwidth]{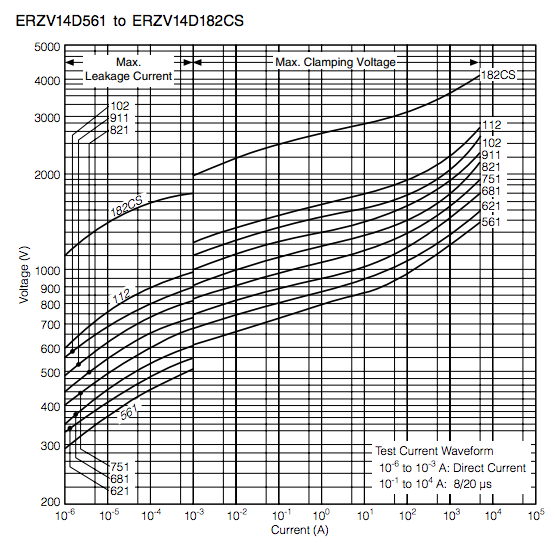}
\includegraphics[width=0.50\textwidth]{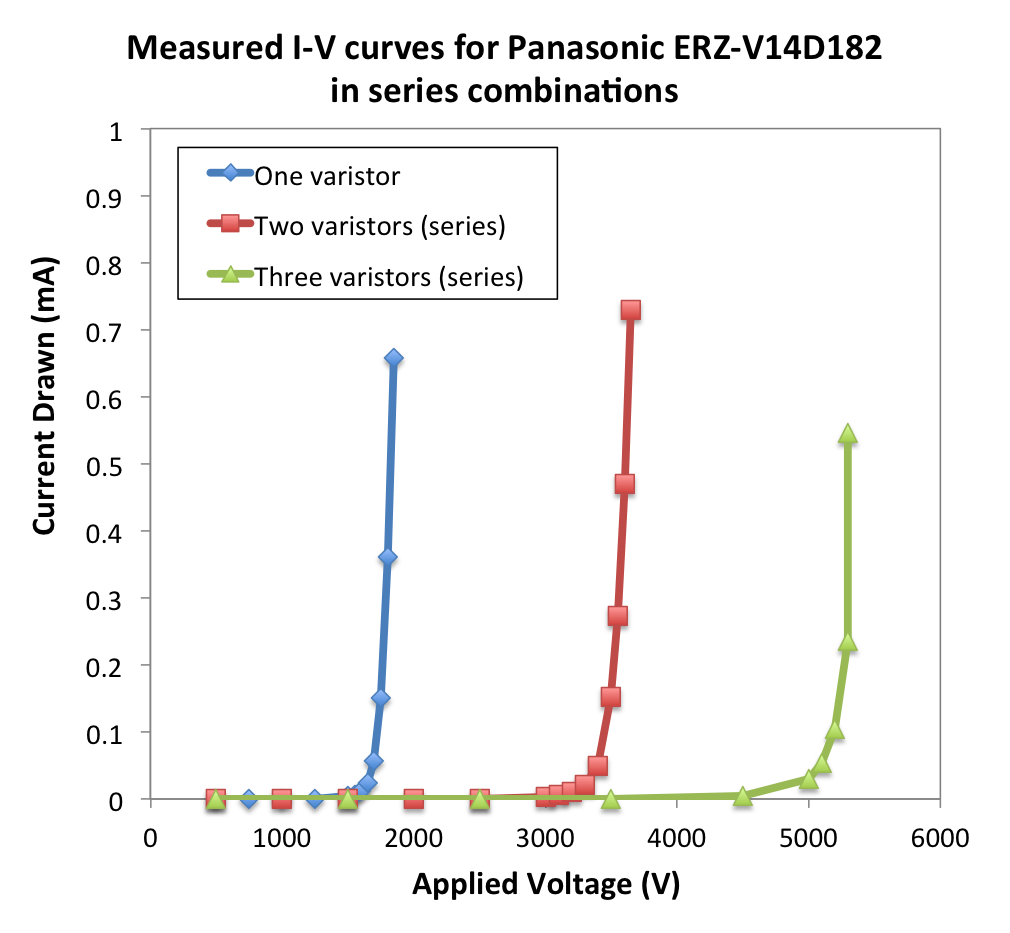}
\caption{Left: I-V characterstic for the Panasonic ERZ range of varistors, from Ref.~\cite{pano}. Right: measured I$-$V characteristic of series combinations of Panasonic ERZ-V14D182 varistors in air, shown on a linear scale to demonstrate additivity of clamping voltages.  The clamping voltage scales by a factor of $N$ for $N$ varistors in series.\label{fig:IVCurves} } 
\end{figure}

Commercially available varistors of the appropriate size for mounting on a MicroBooNE-scale TPC field cage span clamping voltages from 6 to 1800~V.  Since this voltage range is below the required 2~kV, no single device appears to be appropriate for field cage protection, where the nominal applied voltage 2~kV must be held with low current leakage.  This is a commonly encountered situation in the field of surge protection, and can solved by applying $N$ devices in series~\cite{Woodworth}.  The applied voltage is then shared approximately equally between devices, effectively increasing the clamping voltage by a factor of $N$.  Even if there are small mismatches between the I$-$V characteristics of the individual units, the smoothly varying nature of the I$-$V curve ensures that a stable voltage sharing is achieved for any series combination.  The measured I$-$V characteristic for a single varistor and sets of two and three in series are shown in Figure~\ref{fig:IVCurves}, right.

By adjusting the varistor grain size, device dimensions and dopant composition, different clamping voltages and energy tolerances can be achieved.  The device we investigate in this paper is the Panasonic ERZ-V14D182~\cite{pano}, whose I$-$V curve is shown on the left hand side of Figure~\ref{fig:IVCurves}. This device has a clamping voltage of 1800~V, a maximum surge energy rating of 510~J for an industry standard 8/20~$\mu$s pulse, and is rated for the temperature range $-40$ to $+85$~C.  Two or three devices in series can provide an appropriate clamping behavior for the MicroBooNE application.

\subsection*{Gas Discharge Tubes (GDTs)} \label{sec:GDTs}

Gas discharge tube arrestors, or GDTs, are commercially available surge protection devices which consist of a small volume of gas inside a ceramic housing, in contact with two electrodes.  In the event of an over-voltage, a spark forms between the electrodes which allows a large current to flow.  Current continues to flow until the voltage across the device drops below the extinction voltage.  The amount of charge which flows across the device before extinction occurs depends on the available current feeding the discharge.  At the spark voltage, GDTs undergo a discontinuous transition from the resistive to conductive state, causing the device to act as a crowbar.  This transition, and the eventual extinction, may not always occur at exactly the same voltage with repeated pulses.  This means that unlike varistors, GDTs do not have a smooth and well defined I$-$V characteristic.  

By changing the device geometry, and gas composition and pressure, different spark-over voltages can be achieved.  For the application described above, we require a GDT with a DC breakdown voltage above 2~kV, but below 30~kV.  The device we investigate in this paper is the EPCOS A71-H45X~\cite{ep_datasheet}, which has a DC spark-over voltage of 4500~V, an insulation resistance specification of $>10$~G$\Omega$, a current rating of 2.5~kA for an industry standard 8/20~$\mu$s pulse, and a temperature rating of $-40$ to $+90$~C.  Extinction occurs at around 500 V warm and cold.  Devices with both higher and lower DC spark-over voltages are available.  For the model chosen, a single unit can provide the appropriate clamping behaviour for the MicroBooNE applicaiton.

\section{Behaviour under cryogenic conditions} \label{sec:BehaviorInCryogenics}

It is vital for the proper function of a surge arrestor that it remain sufficiently insulating under the nominal applied voltage, and become sufficiently conducting to provide clamping in a surge condition.  We observe that both varistors and GDTs experience changes in their electrical characteristics when operated at cryogenic temperatures.  The varistor voltage and GDT DC spark-over voltage were characterized for 168 varistors and 60 GDTs both in air (warm) and in liquid argon (cold).  Batches of around 20 devices were mounted on a specially constructed wheel which can be steered to place electrical contacts across each component sequentially.  This test wheel was operated both in the air and immersed in an open dewar of liquid argon to characterize a large sample of devices.

\begin{figure}[t]
\centering \includegraphics[width=0.48\textwidth]{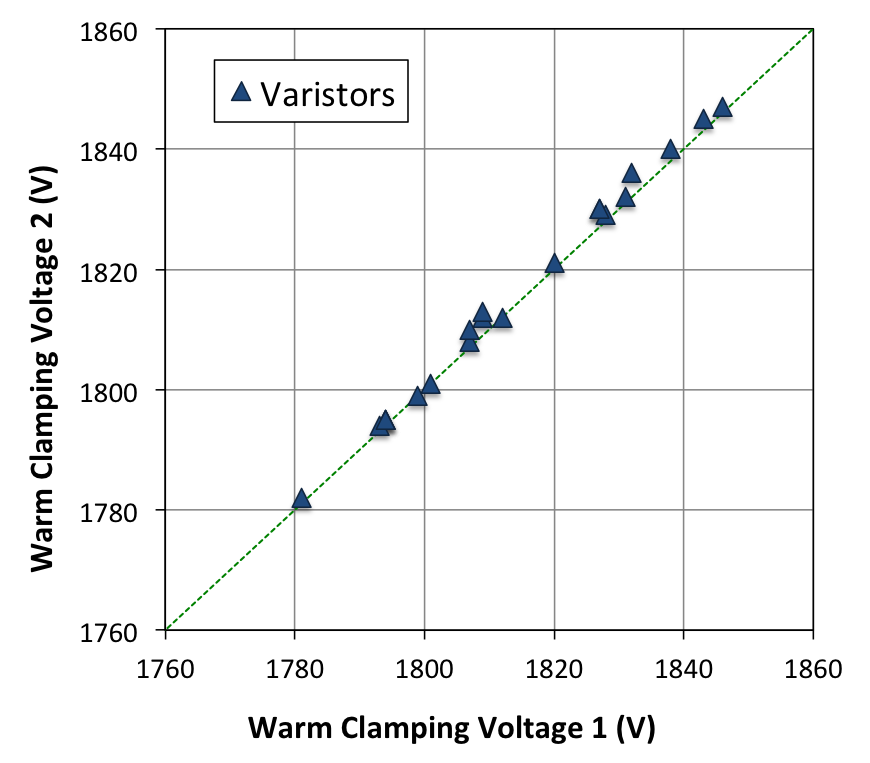}
\includegraphics[width=0.48\textwidth]{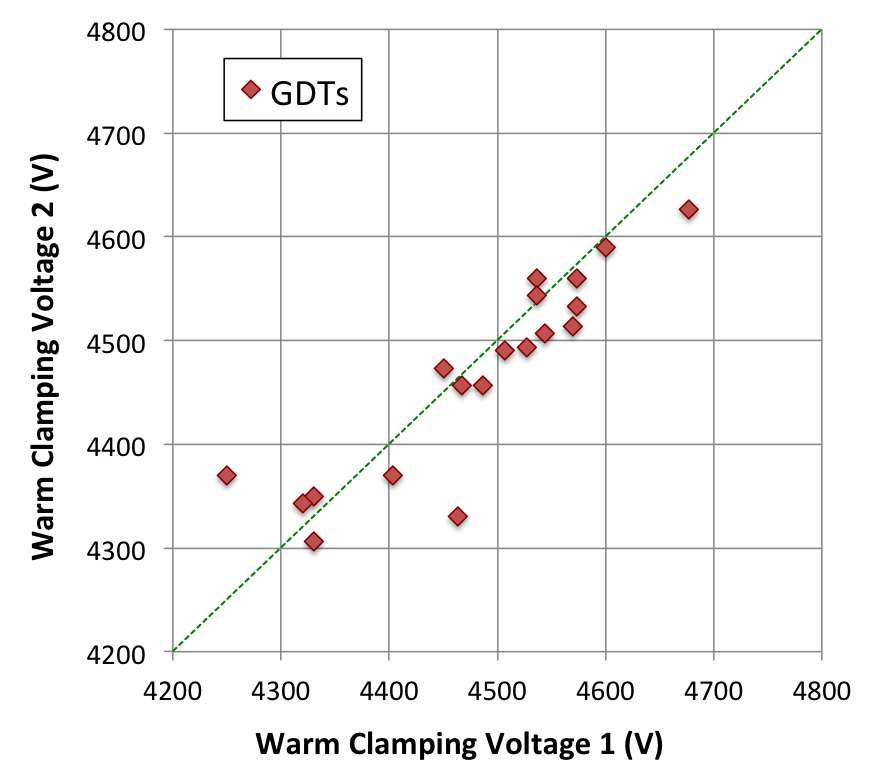}
\caption{Repeatability of warm clamping measurement for varistors (left) and GDTs (right).  Measurement 1 and measurement 2 were made for the same device, approximately 30 minutes apart. The varistors have a very stable clamping point, whereas the voltage at which GDTs clamp is somewhat random each time.\label{fig:WarmRepeatability} }
\end{figure}

For each device, the voltage supplied using a calibrated DC high voltage supply is gradually increased from zero.  For varistors, the voltage supplied when a total current of 0.5~mA is drawn is recorded.  For GDTs, we record the largest voltage supplied before the GDT transitions into a conducting mode, exceeding the 0.8~mA trip current of the power supply.  In what follows we will refer to the relevant recorded voltage as the clamping voltage, despite the different definition in each case.  For the first batch of 20 devices the repeatability of the characterization was established by measuring the warm clamping voltage two times each.  These data are shown in Figure ~\ref{fig:WarmRepeatability}.  The varistors have a clamping voltage which is repeatably measurable to within a few volts. The GDTs have more variability in the clamping voltage, taking a somewhat random value each time.  All devices clamp well within their specification, which is $1800^{+180}_{-100}$~V for the varistors, and $4500 \pm 900$~V for GDTs.

\begin{figure}[t]
\centering \includegraphics[width=0.48\textwidth]{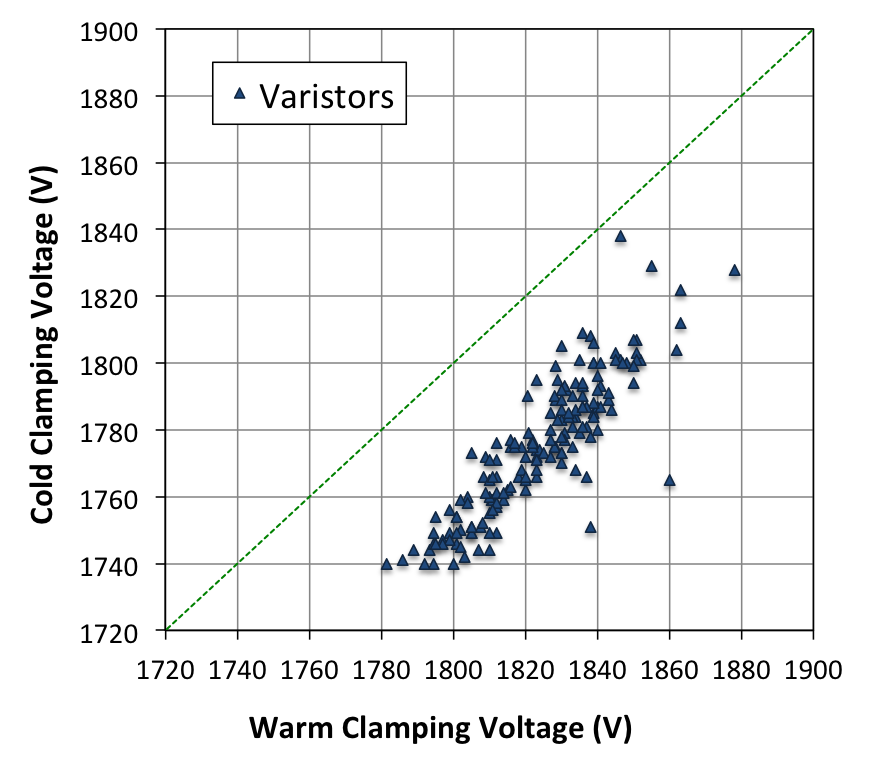}
\includegraphics[width=0.48\textwidth]{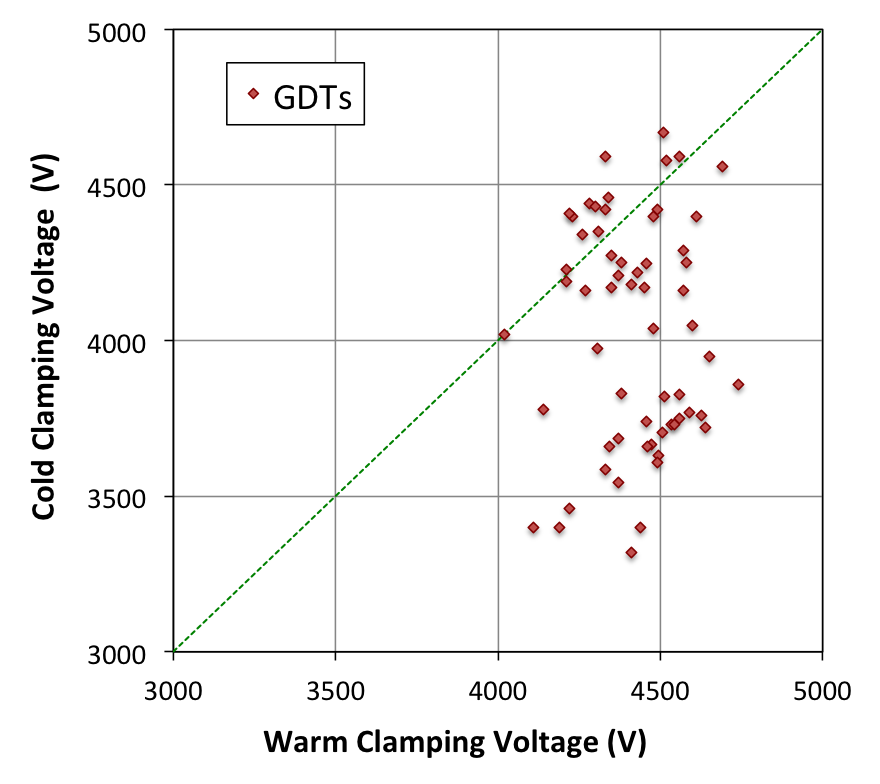}
\caption{Cold versus warm clamping voltage for varistors (left) and GDTs (right).\label{fig:WarmVsCold}} 
\end{figure}

The clamping voltage was compared warm and cold for the two samples.  The cool-down procedure involved holding the devices in the argon vapor around 10~cm from the liquid surface for around 10~minutes, before immersing the wheel completely in the liquid.  A further 10~minutes was allowed for the temperature of the wheel and attached devices to equilibrate with the liquid argon.  The warm and cold clamping points of each device are shown in Figure~\ref{fig:WarmVsCold}.  For varistors the clamping voltage always drops when cold, and there is a strong correlation between the warm and cold values. For GDTs there is no clear correlation, although the cold clamping voltages are typically lower and more scattered than warm.

As well as being suitably conducting in the surge regime, a surge protection device must be sufficiently insulating under the nominal applied voltage.  For our application this is 2~kV for single devices, or 2~kV/$N$ per device for $N$ devices in series, where for varistors $N$ must be at least 2.  The leakage current for both varistors and GDTs was investigated both warm and cold.

\begin{figure}[t]
\centering \includegraphics[width=0.98\textwidth]{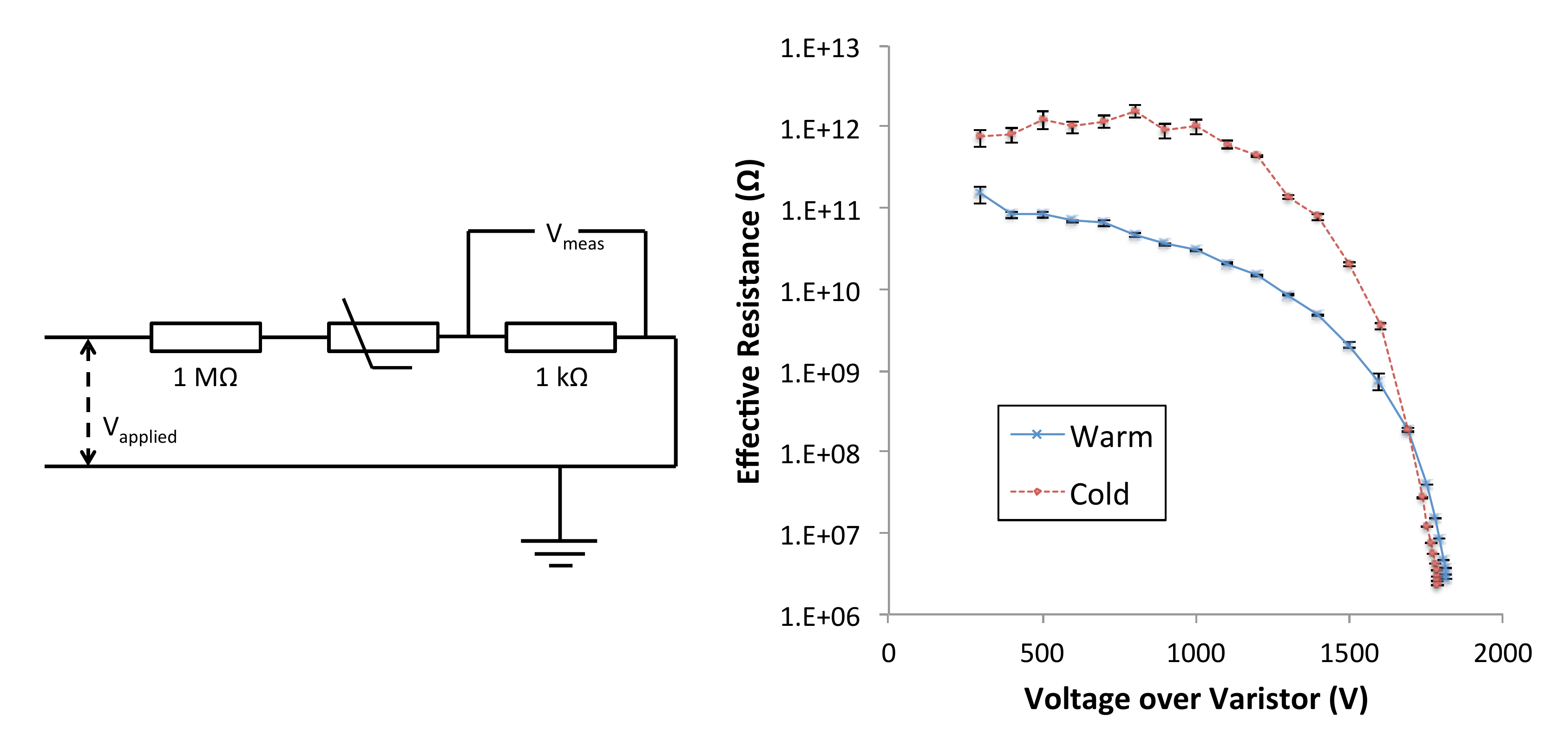}
\caption{Left: circuit used to characterize varistor leakage behavior in warm and cold. Right: effective varistor resistance measured in warm and cold conditions.}
\label{fig:VaristorLeakage} 
\end{figure}

For varistors, a stable DC current is drawn at the nominal applied voltage.  To measure the varistor curve warm and cold, the circuit shown in Figure \ref{fig:VaristorLeakage}, left was used.  The measured I$-$V characteristic warm and cold is shown in Figure~\ref{fig:VaristorLeakage}, right.  The interpretation of the leakage current over long timescales is not trivial, since the varistor has a finite capacitance which charges through a large resistance.  This results in a capacitive component to the leakage current which decays over a few minute timescale as well as a possible DC resistive component.  The measurements shown in figure~\ref{fig:VaristorLeakage} represent the leakage current measured within a few seconds of voltage being applied, and give a lower limit to the effective insulating resistances which can be expected at long times.

\begin{figure}[t]
\centering \includegraphics[width=0.65\textwidth]{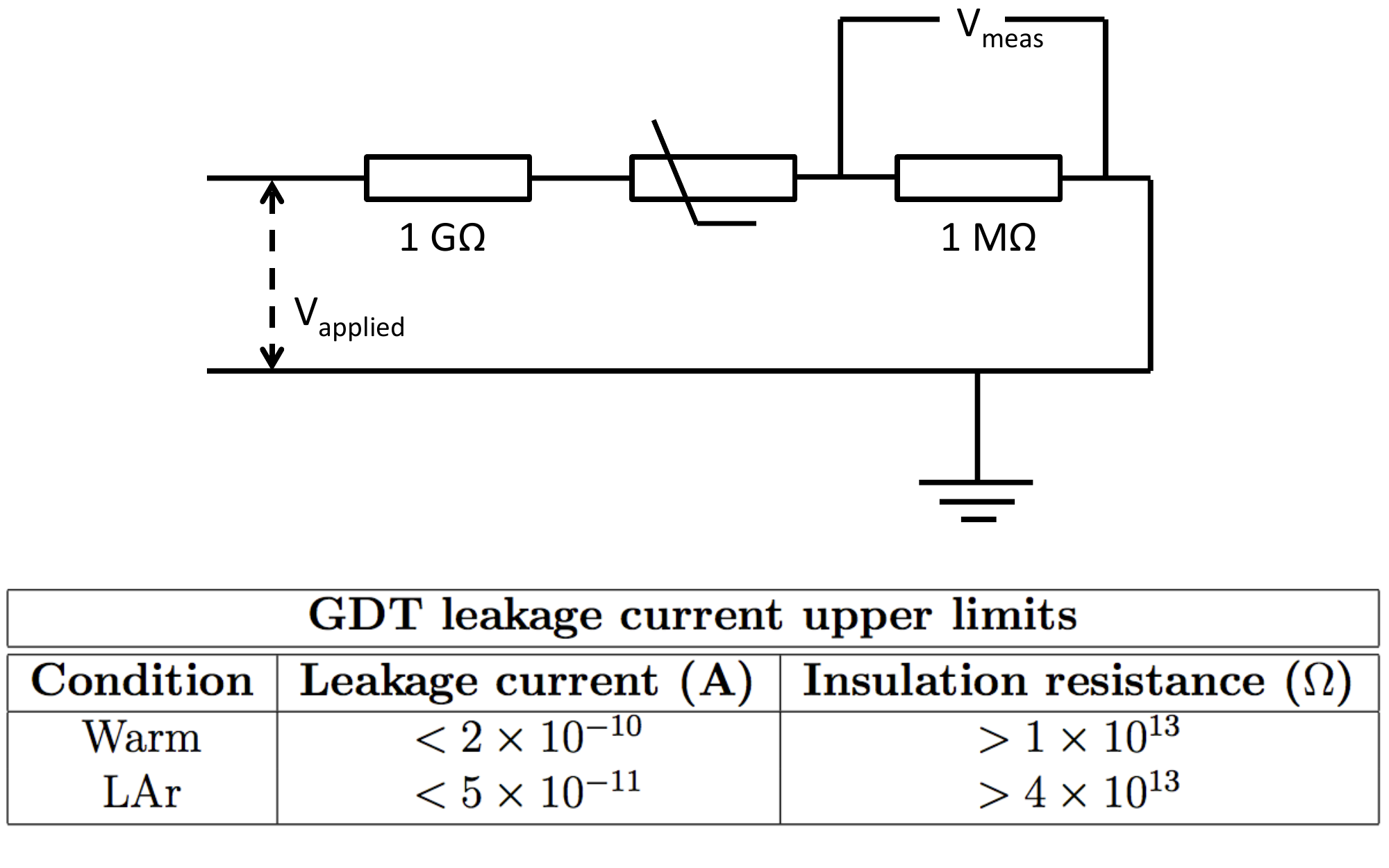}
\caption{Top: circuit used to characterize GDT leakage behavior in warm and cold. Bottom: effective leakage current and resistance measured in warm and cold conditions at 2~kV applied voltage.\label{fig:GDTLeakage} }
\end{figure}

For GDTs, the leakage current was measured using the circuit shown in Figure~\ref{fig:GDTLeakage}, top.  If the resistance of the device is much greater than 1~G$\Omega$, the full applied voltage in this circuit is dropped across the surge arrestor.  By measuring the voltage across the 1~M$\Omega$ resistor, we measure the current flowing in the circuit and hence the insulating resistance of the surge arrestor.  When the circuit was biased at 2~kV, a small, fluctuating voltage was observed.  We record the largest value of the fluctuating voltage to give a lower limit on the leakage current, which is tabulated in Figure~\ref{fig:GDTLeakage}.  

For the MicroBooNE application, an insulating resistance much larger than the $2.5 \times 10^8~\Omega$ field cage resistors is required.  Both devices satisfy this criterion by a large margin.

Another requirement is that the devices must operate stably in a cryogenic environment over long timescales without experiencing random discharges, large leakage fluctuations or damage caused by argon seepage.  To address the first two points, 14 of each device were biased at the nominal operating voltage in parallel, in an open liquid-argon filled dewar.  The trip on the power supply was set to 100~nA.  For both GDTs and varistors, the power supply remained un-tripped for over 72~hours, demonstrating 1000~device-hours of stable operation. 

A longer term test was performed for two of each device over three weeks in a high purity argon environment.  This tests not only that the devices do not degrade when submerged in argon, but also that they operate as expected in the dielectric environment of high purity argon. All four devices continued to exhibit the expected behavior in the insulating and clamping regimes for the full three weeks of the study.

\begin{figure}[t]
\centering \includegraphics[width=0.48\textwidth]{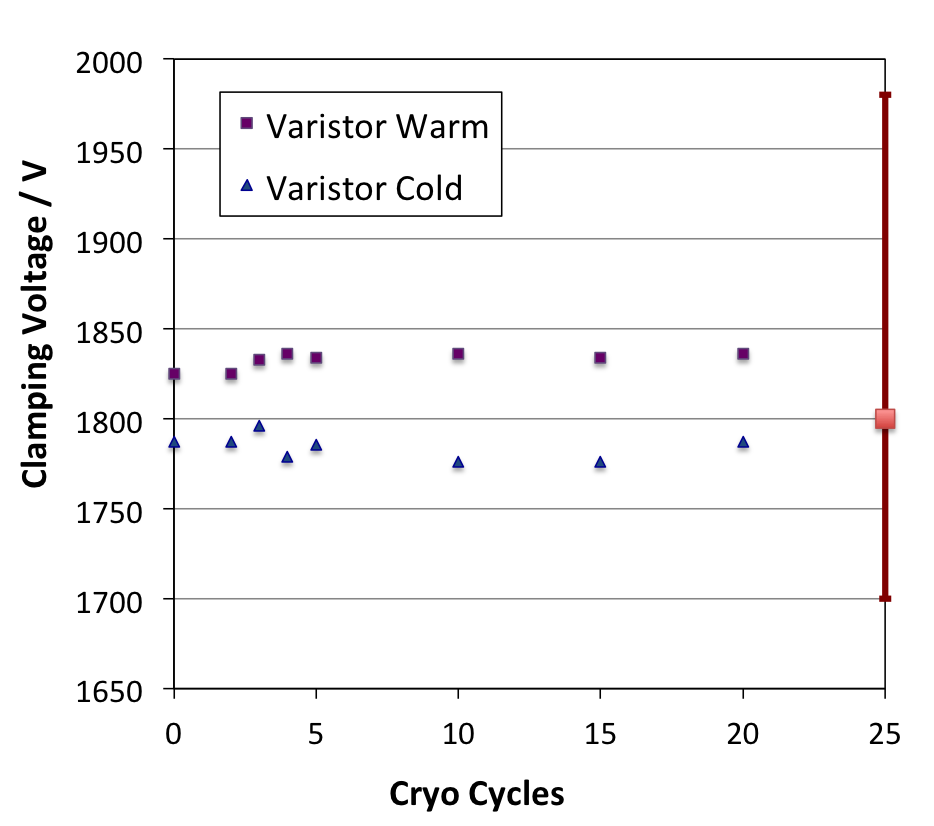}
\includegraphics[width=0.48\textwidth]{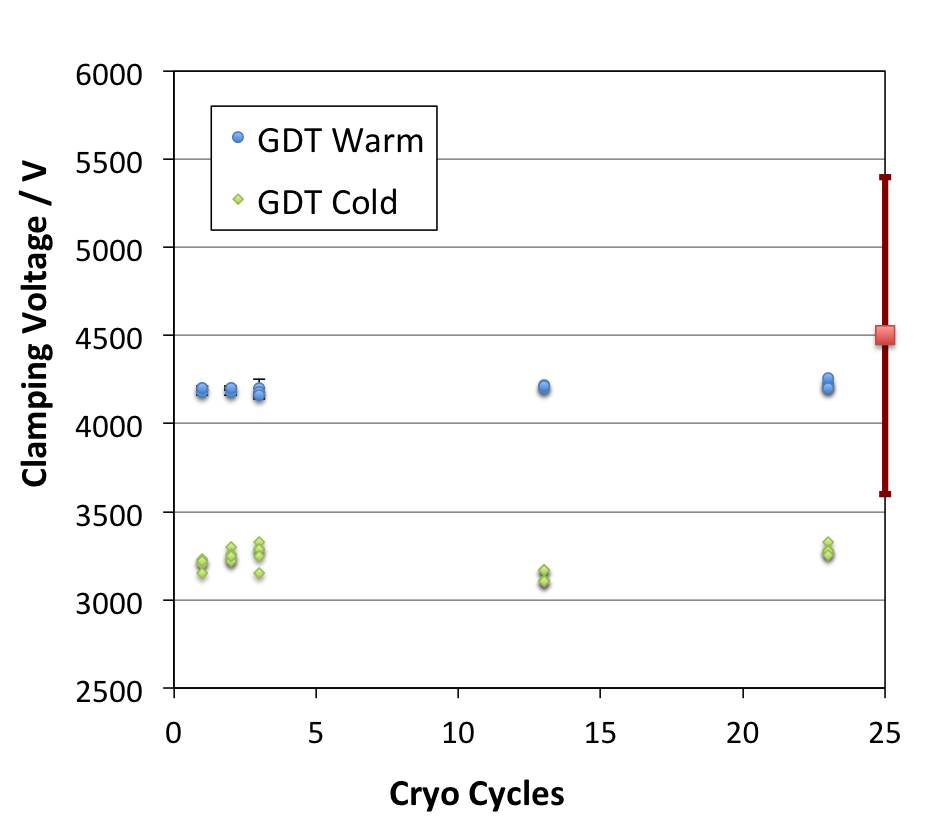}
\caption{The effect of cryo-cycling upon varistor and GDT clamping voltage.  Both devices are electrically unaffected by repeated cycles.  The red bar on the right of each plot shows the specification provided by the manufacturer for the warm clamping voltage.\label{fig:CryoStrength} }
\end{figure}

In addition to being submerged in liquid argon, a batch of GDTs and varistors were also
tested by submerging them in liquid nitrogen in an open dewar.  The devices display similar insulating and clamping properties in this environment as they do in liquid argon. This is possible for the GDTs even though they contain nitrogen gas as the spark medium, because the gas is held at a sufficiently low pressure that it is not allowed to liquefy within the device.  This point will be discussed further in Section~\ref{sec:Purity}.

Although repeated or fast transition between liquid argon temperatures and ambient
temperatures is never foreseen for devices used in LArTPC's, the effect of such
cryogenic cycling was investigated for these devices. These tests were done to
demonstrate both their mechanical and electrical robustness under extreme cryogenic
conditions.  GDTs were observed to be very robust - the device was mechanically and electrically unaffected by 23 violent cycles between air and liquid argon environments, being heated with a heat gun between each dunk.  The warm and cold clamping voltages at various times during this process are shown in Figure \ref{fig:CryoStrength}, right.  Varistors were not quite so robust, requiring a gentler cool-down procedure.  Under the more violent procedure applied to GDTs, some varistors were observed to develop small hairline cracks in their epoxy coating.  These cracks did not appear to affect the function of the device, having no impact upon its clamping or insulating properties.  With a gentler cycle more closely resembling the conditions expected during a detector fill, with the varistor held briefly in the vapor between each immersion or removal, there was no damage to the component either visible by eye or electrically.  The warm and cold clamping voltages at various stages during this process are shown in Figure \ref{fig:CryoStrength}, left.

\section{Robustness under repeated surges in liquid argon} \label{sec:RobustnessUnderSurges}

\begin{figure}[t]
\begin{centering}
\includegraphics[width=0.8\columnwidth]{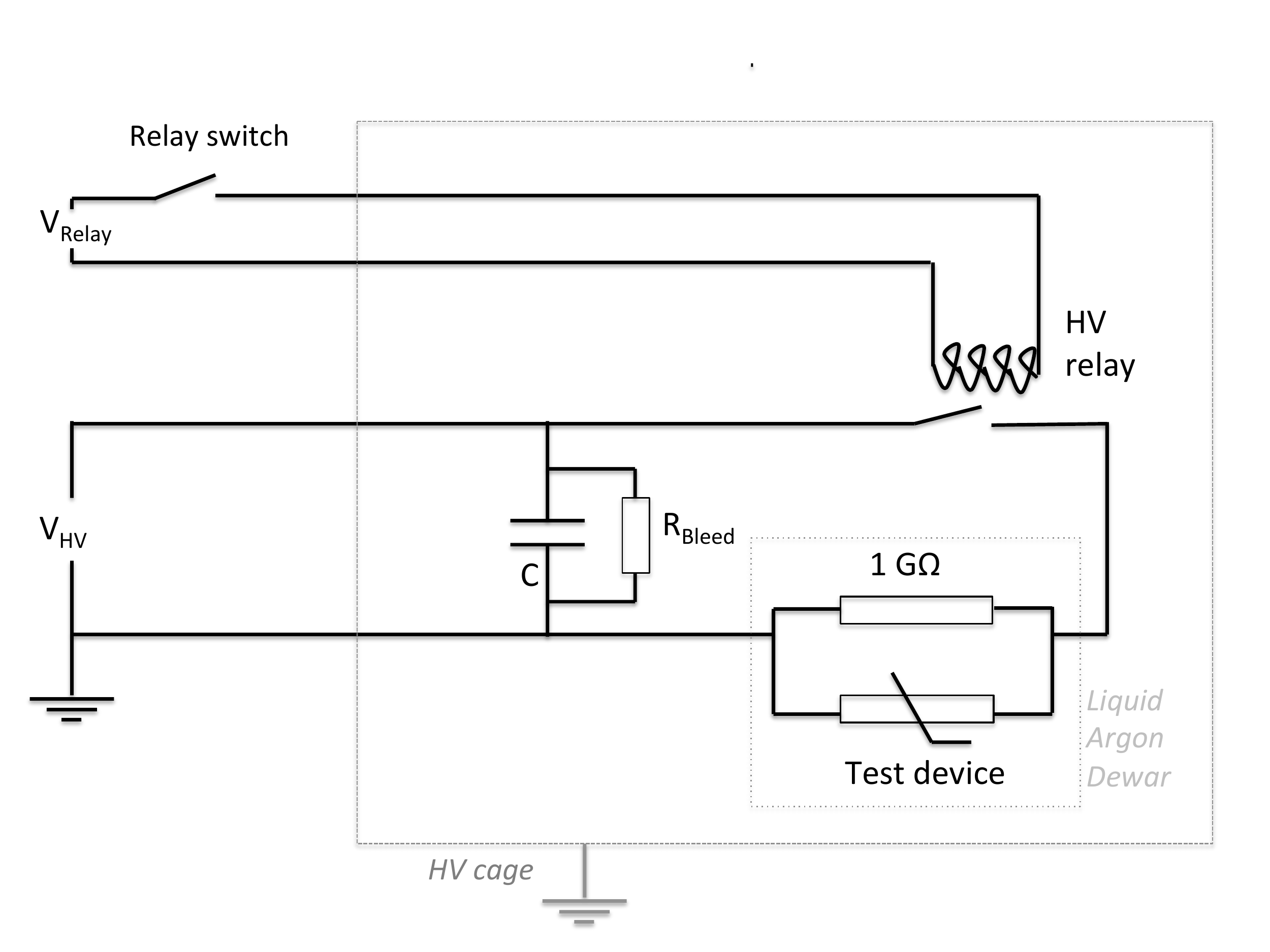}
\par\end{centering}

\caption{Circuit diagram of the HV pulse test.  The rate of discharge into the test device is limited only by the resistance of the cables in the system. \label{fig:BeastCircuitDiagram}}
\end{figure}

An important characteristic of a surge protection device is its lifetime under the influence of repeated surges.  
All surge protection devices have a finite lifetime and ratings.  Where given, these usually relate to an industry standard 
surge condition such as an 8/20~$\mu$s or 10/1000~$\mu$s (rise time / fall time) current pulse, delivered in air and at room temperature.  

The expected use cases in LArTPCs do not reflect these idealized conditions.  Furthermore, in many industrial 
applications, surge protection components can be considered as sacrificial, being replaced after the end of their 
lifetime.  In cryogenic experiments, the experiment design may make them inaccessible for long intervals, during 
which time they must survive repeated surges.  Understanding the longevity of these devices in these conditions is of central 
importance.

\begin{figure}[t]
\centering \includegraphics[width=0.48\textwidth]{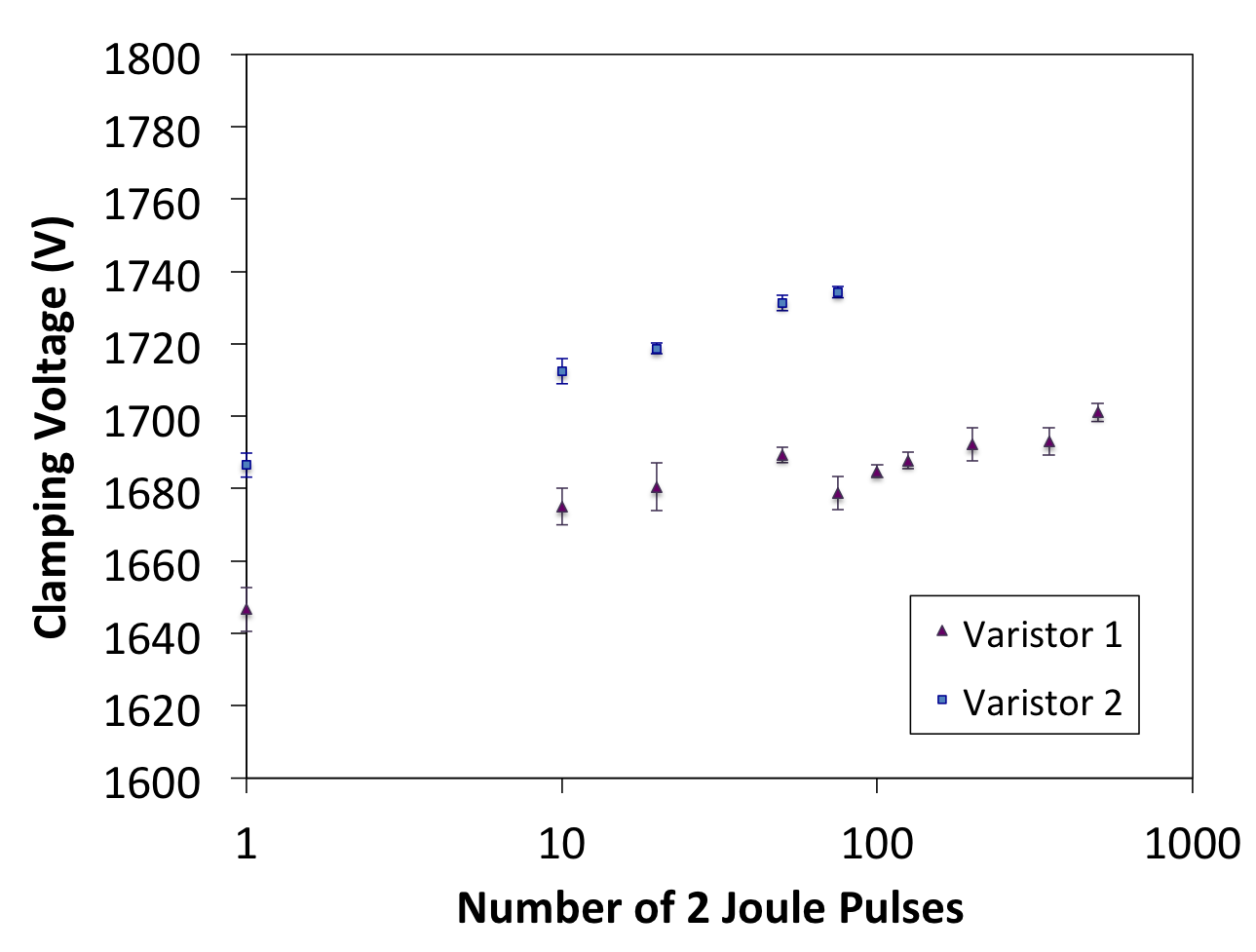}
\includegraphics[width=0.48\textwidth]{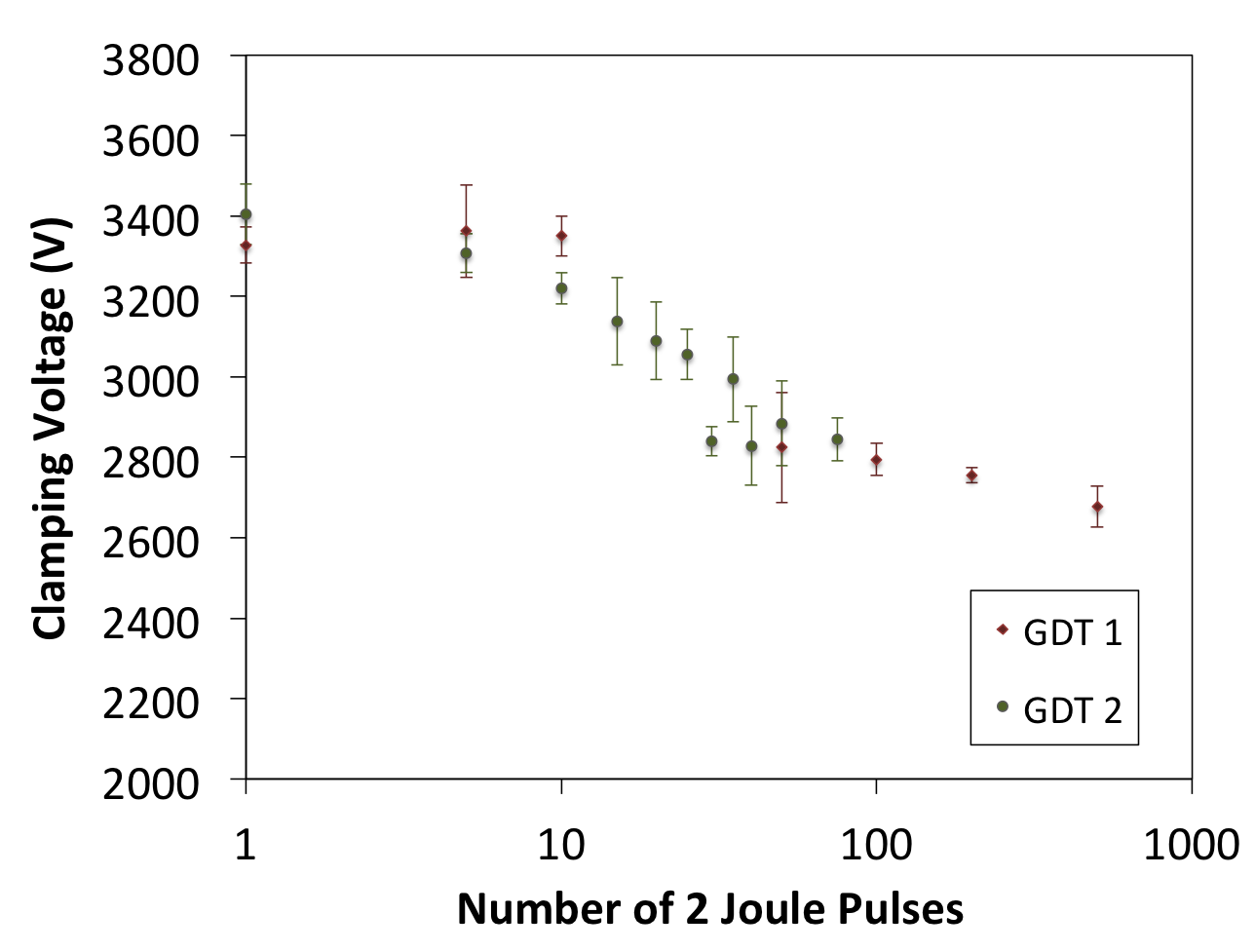}
\centering \includegraphics[width=0.48\textwidth]{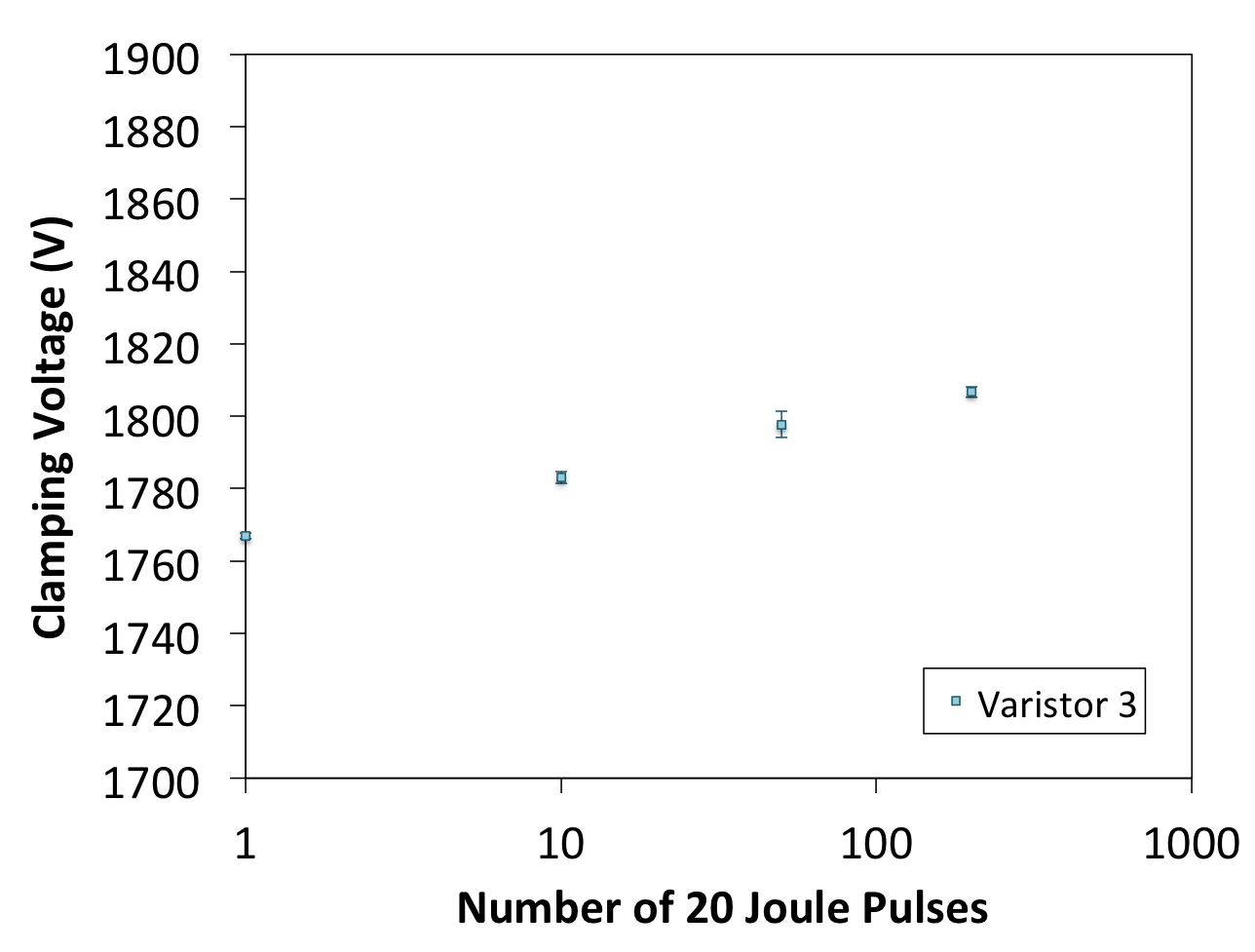}
\includegraphics[width=0.48\textwidth]{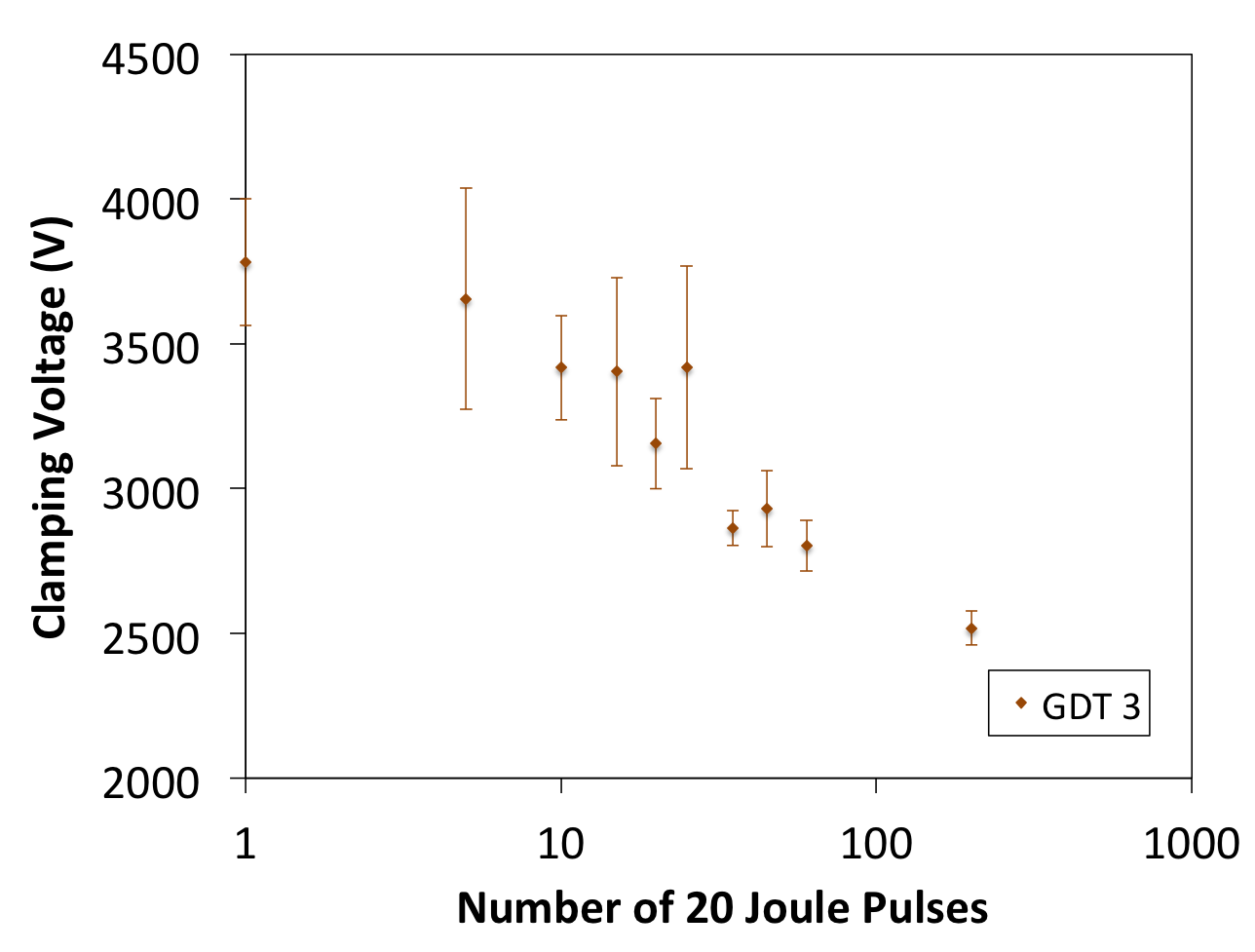}
\caption{Change in clamping voltage of varistors (left) and GDT's (right) with repeated 2J and 20J surges in liquid argon. \label{fig:ClampingSurges} }
\end{figure}

To address these questions for GDTs and varistors a test stand was designed to deliver voltage surges of the scale expected in a HV discharge event in a large LArTPC.  In MicroBooNE, the largest expected discharges would deposit 2~J into a device in a worst case scenario.  To test the robustness of the devices, controlled voltage surges of known energy were produced charging a large capacitor to a known voltage using a high voltage supply, and then rapidly discharging 
the capacitor through the device under test, in an open dewar of liquid argon. The discharge is triggered by closing a high voltage relay. A 2~J pulse is generated by charging a 50~nF capacitor to 9~kV and discharging it through the capacitor. 20~J pulses were also delivered to over-test the device, generated by charging a 1~$\mu$F capacitor to 6.5~kV.  A schematic of the test apparatus is shown in Figure~\ref{fig:BeastCircuitDiagram}. 

Using this apparatus, 500 pulses of 2~J were delivered to one GDT and one varistor, each submerged in liquid argon.  At various intervals during the application of the 500 test pulses the clamping voltage for the device under test was remeasured, as shown in the top two plots of Figure~\ref{fig:ClampingSurges}.  Both devices experienced some drift in the clamping voltage.  The varistor clamping voltage rose within the first few pulses before stabilizing, but always remained within the specifications for the device.  This is in contrast to the degradation behavior reported in the literature for varistors, which suggests that the clamping voltage experiences a steady decay towards zero~\cite{var_material_science}.  However, modern varistors are known to have different degradation behavior to those available at the time of those sources due to the introduction of new dopants and manufacturing processes \cite{Woodworth}.  The GDT clamping voltage is observed to drop with repeated surges, which is consistent with the reported degradation behavior in the literature~\cite{GDTBourns}.  The approximately exponential decay suggests that the device will remain above the critical 2~kV threshold for many more than 500 surges.  A second specimen of each device was subjected to 75~pulses to check the reproducibility of the degradation behavior, and a consistent behavior was observed.  With pulses of 20~J, a similar trend is observed for both varistors and GDTs. For varistors the change in clamping voltage is not appreciably different from the 2~J case.  For GDTs the degradation is more dramatic, but the device still appears robust for up to 200~pulses. These data are shown in Figure~\ref{fig:ClampingSurges}, bottom.

\begin{figure}[t]
\centering \includegraphics[width=0.48\textwidth]{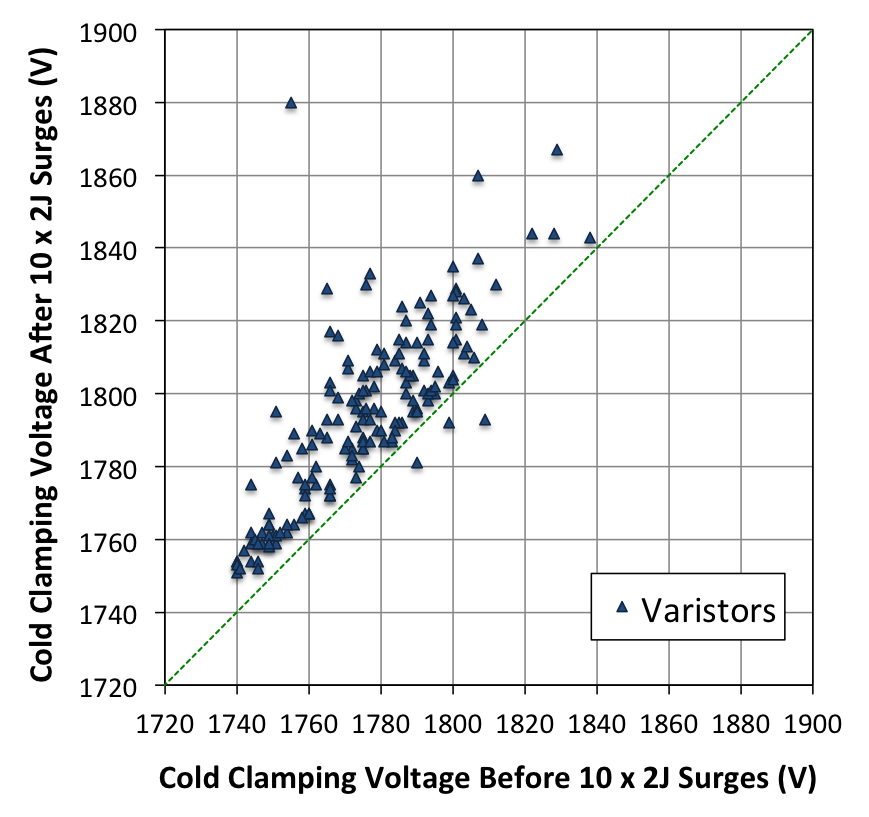}
\includegraphics[width=0.48\textwidth]{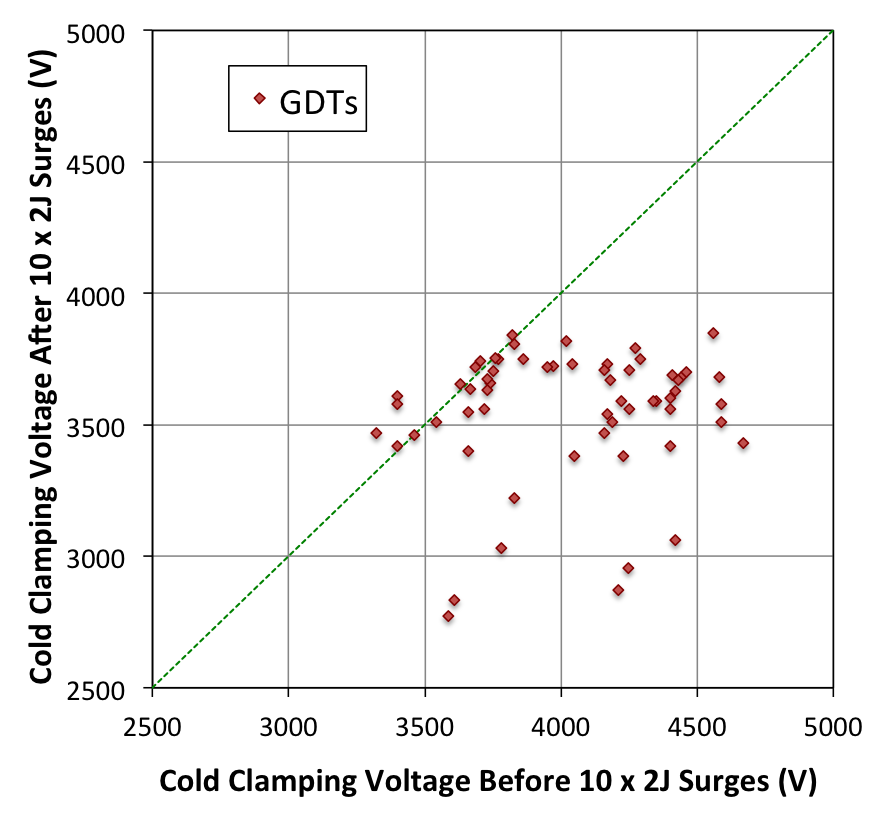}
\caption{Shift in clamping voltage for 168 varistors and 60 GDTs with ten 2~J surges. \label{fig:HighStatsSurges} }
\end{figure}

Using the wheel apparatus described in Section~\ref{sec:BehaviorInCryogenics}, the 168 varistors and 60 GDTs were subjected to ten pulses of 2~J each, administered using a 100~pF capacitor charged to 6.3~kV.  The distribution of clamping voltages before and after these pulses are shown in Figure~\ref{fig:HighStatsSurges}.

The tests described above exercise the energy dissipation capability of each surge arrestor.  In these tests, a pulse of voltage significantly less than the design high voltage of a large LArTPC was used.  This is reasonable, because the surge arrestors prevent the potential differences in the system from rising above the clamping voltage, so they should not have to stand off any voltage significantly above this clamping voltage during a discharge event.  In spite of this, there remains a possibility that rapid transient over-voltages up to the full supply voltage may be encountered on short timescales due to inductive effects in the system.  To test that the devices are robust against such over-voltages, and offer suitable protection to vulnerable components in this condition, a 150~kV transient test was performed.

\begin{figure}[t]
\centering \includegraphics[width=0.8\textwidth]{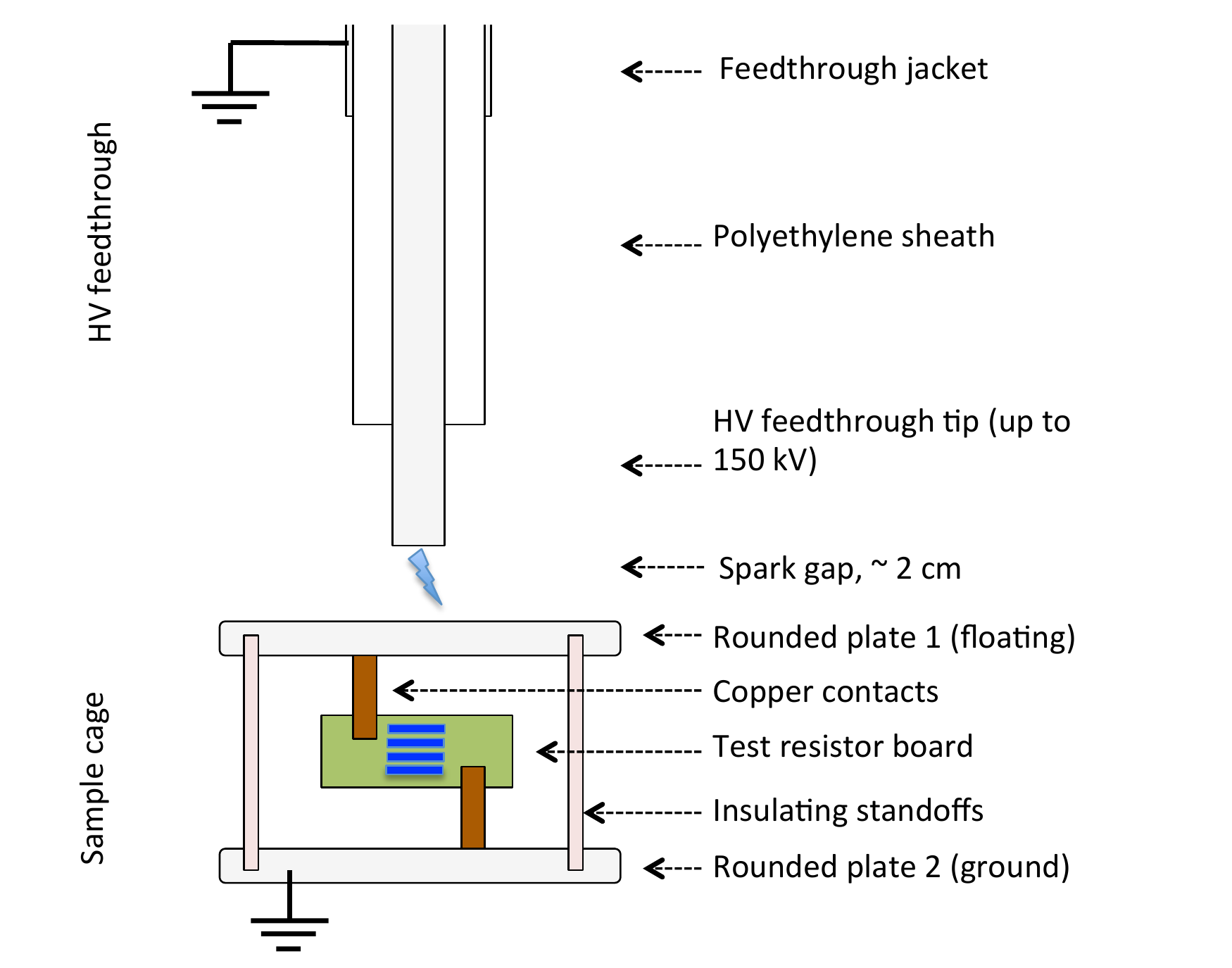}
\caption{Diagram of the apparatus used to perform the 150~kV transient test.\label{fig:TransientTestDiagram} }
\end{figure}

\begin{figure}[t]
\centering \includegraphics[width=0.48\textwidth]{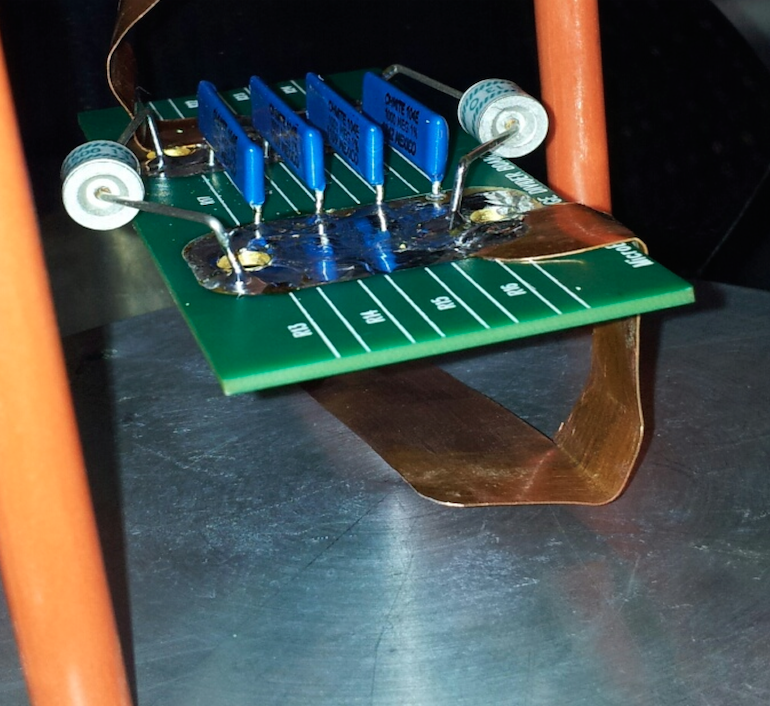}
\includegraphics[width=0.463\textwidth]{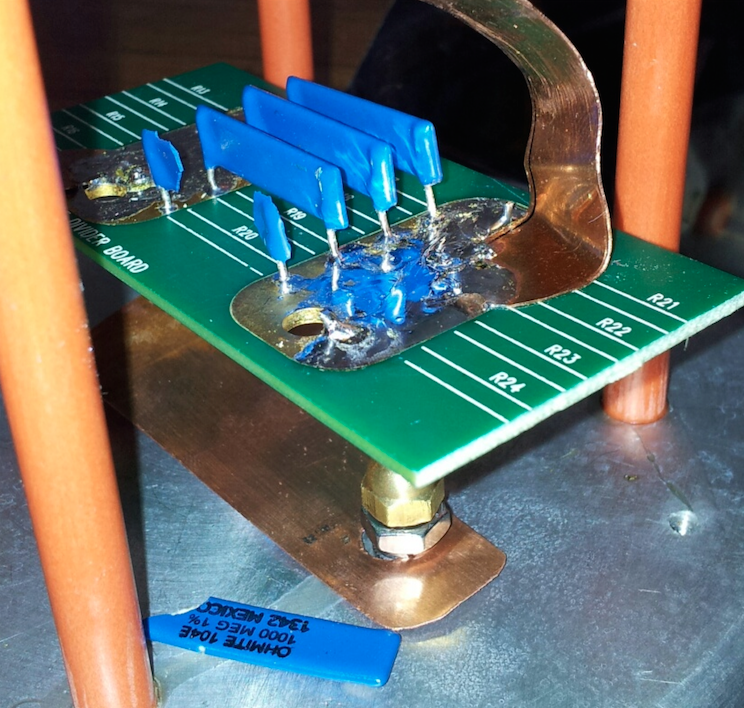}
\caption{Left : Photograph showing the protected resistor board following the delivery of $20\times150$~kV transient surges.  Right : photograph showing the same board after delivery of a single 70~kV transient surge when no surge protection is applied.\label{fig:TransientTestPhotos} }
\end{figure}

The test setup is shown in Figure~\ref{fig:TransientTestDiagram}.  One MicroBooNE resistor board, an example of which is shown in Figure \ref{fig:TransientTestPhotos}, complete with four Ohmite Slim-Mox 104E
1~G$\Omega$ resistors was mounted inside a sample cage consisting of rounded metal plates separated by plastic rods.  The resistor board with no surge arrestors attached is referred to as unprotected.  The resistors can also be protected by adding surge arrestors in two configurations: 1) two varistors
linked in series and applied in parallel across the resistors and 2) two GDTs applied in
parallel across the resistors.  An example of a protected resistor board mounted in the test cage is shown in Figure~\ref{fig:TransientTestPhotos}(left), with two GDTs in parallel with the four SlimMox resistors.    The cage was immersed in a 200~L open liquid argon cryostat, which was grounded within a high voltage cage.  The bottom plate of the assembly was placed in contact with the lower surface of the cryostat, grounding it.  The top plate remains electrically floating, isolated from ground by the field cage resistors. 

A high voltage feedthrough was inserted into the dewar, with the tip held around 2~cm from the top plate to create a spark gap in the liquid argon.  This feedthrough was charged up to a maximum of 150~kV using a Glassman (series LX) high voltage supply, and held at voltage until a spark developed across the argon gap between the feedthrough tip and the top plate.  At the instant of the spark, a conducting path forms between the feedthrough and the top plate, bringing it up to the full applied voltage.  In the case where a surge protector is applied, the charge stored in the feedthrough and cable, which has a capacitance of around 500~pF, is drained through the spark and the conducting surge arrestor to ground.  This current draw causes the supply to trip.  The total stored capacitive energy in the feedthrough and cable is around 5 J, and a large but not precisely known fraction of this will be dissipated into the surge arrestor. In the case where an unprotected resistor board is used, the resistors have to stand off the full supply voltage, and if this voltage is large enough, one or more of the resistors will fail.  In both cases, the design of the test produces conditions which resemble the expected transient over-voltage condition in a LArTPC spark event.

Sparks began to be observed at 50~kV, and the applied voltage was stepped up in 5~kV steps with at least 20 sparks being delivered at each voltage point, up to the supply limit of 150~kV.  The varistor-protected board was tested first, and after the end of the test the board was removed and the varistors were taken off.  The varistors were characterized and found to display the expected clamping behavior, and the resistors on the board were undamaged.  The test was repeated with GDTs, and again, all components visibly survived the application of sparks up to 150~kV, and the GDTs and resistors exhibited the correct properties after removal.  Finally the GDTs were removed leaving only bare resistors. One of the unprotected resistors failed upon the first spark at a voltage of 70~kV, being destroyed as shown in Figure~\ref{fig:TransientTestPhotos}, right.

This test not only shows that GDTs and varistors can survive brief transients of the full supply voltage, but also provides an explicit demonstration of the protection offered to vulnerable components in the event of a HV surge.  At least one of the same resistors which were able to survive multiple 150 kV discharges when protected with surge arrestors was found to fail at much lower voltages when unprotected.

\section{Practical considerations for use in LArTPCs} \label{sec:PracticalConsiderations}

As well as demonstrating that varistors and GDTs exhibit the required surge protection behavior in a cryogenic environment, it is also important to ensure that they do not interfere with the normal operation of a liquid argon detector in other ways.  In particular, liquid argon TPCs have strict purity requirements.  Also, large scale LArTPCs usually have sensitive optical systems, so the emission of a large photon flux during normal operation would be unacceptable. In this section we describe tests performed to establish possible purity and light emission effects of varistors and GDTs.

\subsection{Purity considerations} \label{sec:Purity}

Two GDTs and two varistors were analyzed with the Fermilab Materials Test Stand (MTS)~\cite{Rebel:2011zzb}.  The MTS consists of a sample cage into which components and materials can be placed, which is lowered into a 250~L volume of high purity liquid argon.  The free electron lifetime in this argon volume is continuously monitored using an electron-drift purity monitor.  After 55~hours in the liquid with the argon filters disabled, the electron lifetime remained at or above 9~ms at all times, in the presence of both GDTs and varistors.  This demonstrates that these devices do not introduce electron-scavenging impurities into LArTPC detectors.

\begin{figure}
\centering \includegraphics[width=0.8\textwidth]{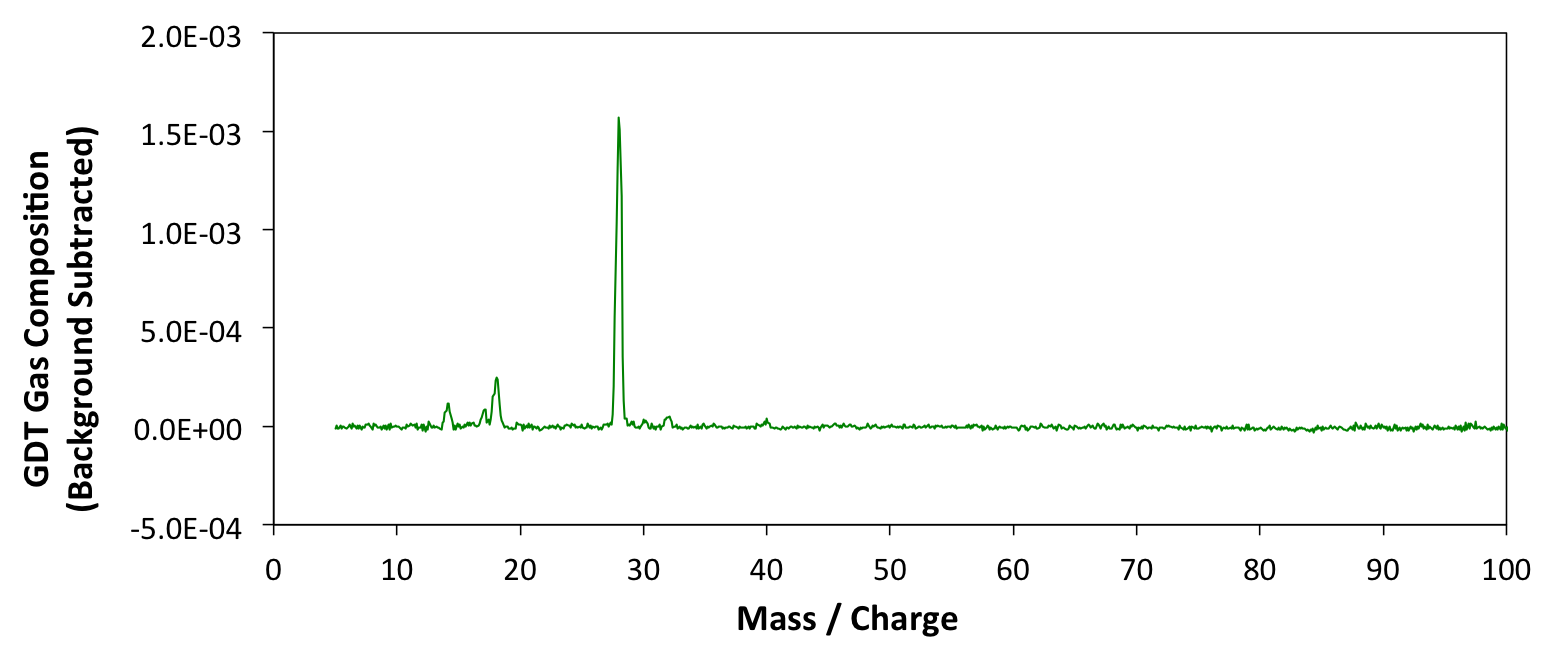}
 \includegraphics[width=0.8\textwidth]{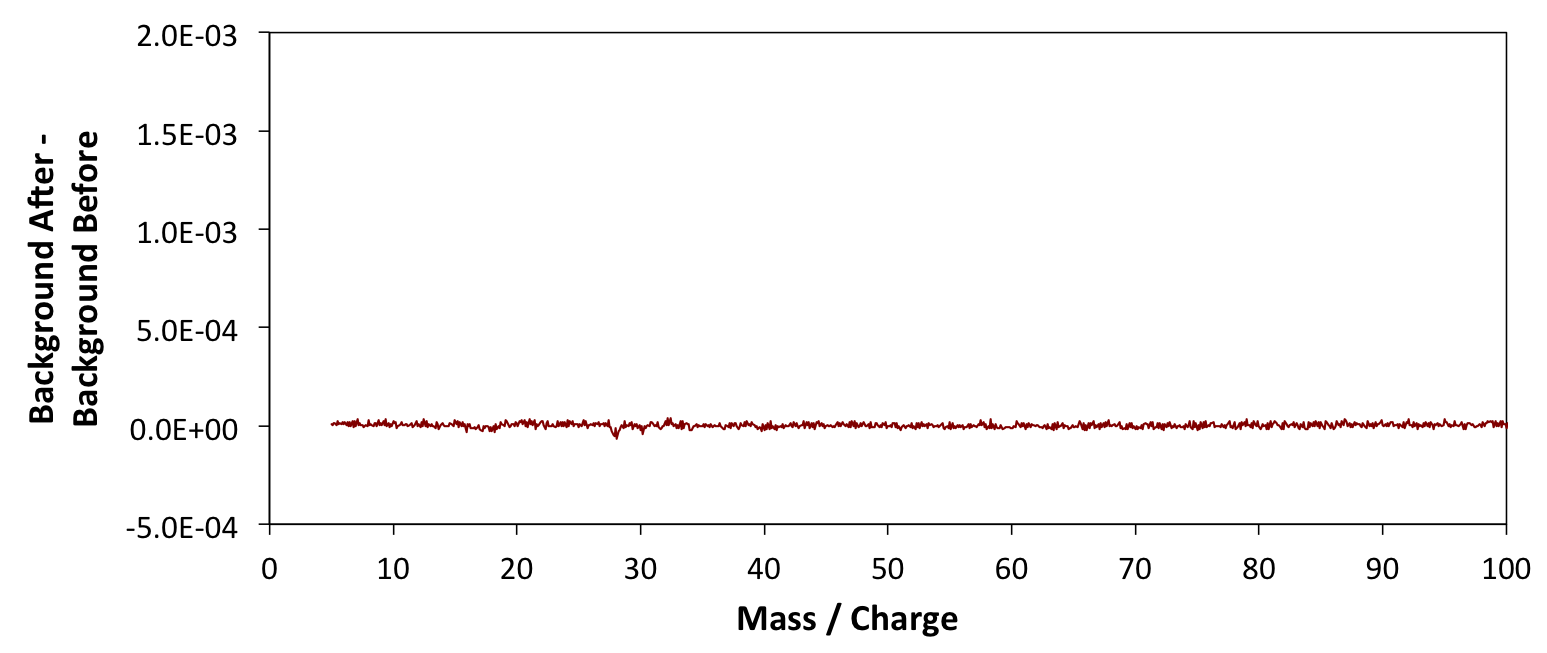}
\caption{Background subtracted UGA spectra.  Top : GDT gas composition. The peaks at 28, 17-18, and 14 correspond to singly ionized nitrogen, water, and doubly ionized nitrogen, respectively. Bottom : repeated carrier gas background measurement with new gas sample in chamber.\label{fig:BGSubtractedSpectra} } 
\end{figure}

\begin{figure}
\centering \includegraphics[width=0.7\textwidth]{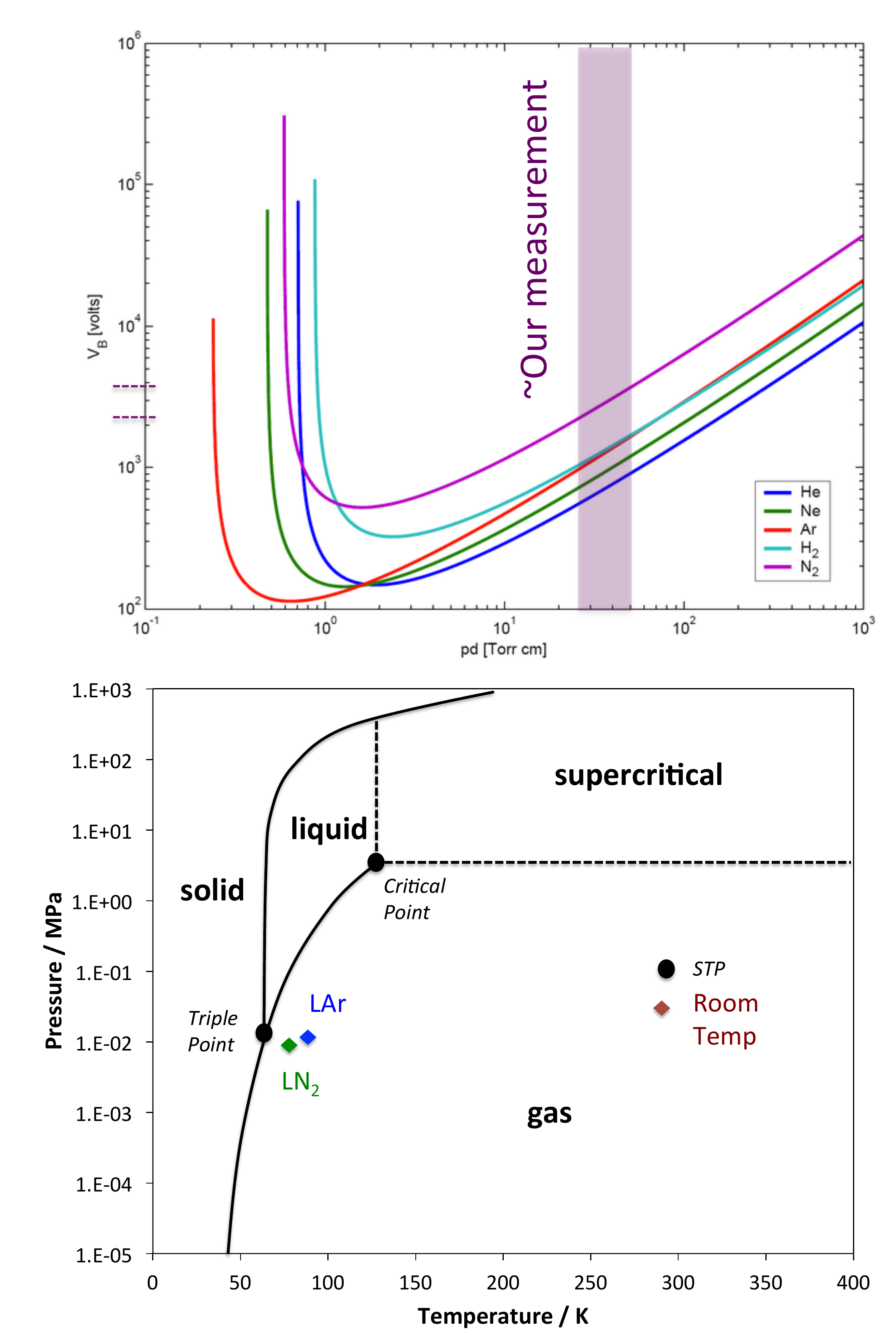}
\caption{Top: Paschen curves for common gases, reproduced from \cite{PaschenCurves}.  Our measurement of the nitrogen pressure in the EPCOS GDT suggests a breakdown voltage of $3-4$~kV, in line with observed behavior.  Bottom: Phase diagram for nitrogen, reproduced from \cite{WolframAlpha}. At the measured concentration and constant volume, nitrogen would be in the pure vapor phase in LAr, LN2, and warm environments, explaining the observed functionality of the GDT.\label{fig:PaschenAndPhaseDiagram} } 
\end{figure}

Varistors' all-solid construction means that even in the event of a catastrophic mechanical failure, the surge protector will bring no new materials into contact with the surrounding medium.  For gas discharge tubes, a rupturing of the device within the detector could lead to the contents becoming dissolved into the argon volume in the event of a catastrophic failure.  The effect of this type of failure upon detector purity was also established.  To determine the composition and concentration of gasses inside the EPCOS A71-H45X, a single unit was broken in the presence of a pre-characterized helium carrier gas.  This mixture was fed to a universal gas analyzer (UGA) \cite{UGARef}, and the mass spectrum of the carrier gas before and after rupture of the GDT were compared.  The difference is shown in Figure~\ref{fig:BGSubtractedSpectra}, top.  After this characterization, the system was flushed with fresh carrier gas, which was again characterized with the UGA.  The stability of the carrier gas composition measurement is shown in Figure~\ref{fig:BGSubtractedSpectra}, bottom, on the same scale.  Using this method, it was established that the GDT contains nitrogen gas at a pressure of 250~Torr.  No other compounds in the mass range $4 < A < 300$ were observed above the noise floor of the UGA, which corresponds to a limit of 3~Torr in the GDT.  No oxygen peak was observed, demonstrating that the measured nitrogen peak is not a result of environmental contamination.  A small peak from water contamination was observed, which is likely caused by outgassing from the steel piping of the analyzer rather than the GDT itself.   Comparison with the Paschen curve for nitrogen confirms that pure nitrogen at the measured pressure, within the geometry of the device, would produce the expected spark breakdown strength, as shown in Figure~\ref{fig:PaschenAndPhaseDiagram}, top.  Comparison with the phase diagram of nitrogen demonstrates that nitrogen at this pressure at room temperature is expected to remain in the vapor phase at both liquid argon and liquid nitrogen temperatures when held at constant volume, which is consistent with our observations of the cryogenic functionality of the device, shown in Figure~\ref{fig:PaschenAndPhaseDiagram}, bottom.

Small contaminations of nitrogen are known to be acceptable for TPC operation.  However, nitrogen at the part-per-million level can have detrimental effects upon liquid argon scintillation light collection~\cite{Acciarri:2008kv, Jones:2013xy}.  Our measurements show that a single GDT, which has a volume of approximately 0.4 cm$^3$, contains $6\times10^{-6}$~mol of nitrogen gas.  As an example, if one GDT were to leak into the MicroBooNE detector, which contains 170 tons of liquid argon, this would introduce a nitrogen contamination of around $1.4\times10^{-12}$. This is far below nitrogen concentrations which are problematic for scintillation light collection.

\subsection{Light emission}

The devices under consideration are to be operated at a nominal voltage of 2~kV (this voltage would be applied across a single GDT, or two or more varistors in series).  In past detectors, some electronic components have been reported to glow when HV is applied~\cite{DChooz}.  In order to establish whether there is light emission associated with the operation of GDTs and varistors at nominal voltages, two of each were installed alongside tests of prototype optical detectors for the LBNE experiment~\cite{Baptista:2012bf}, in the TallBo optical test cryostat at Fermilab~\cite{Bo}.

\begin{figure}
\begin{centering}
\includegraphics[width=0.7\columnwidth]{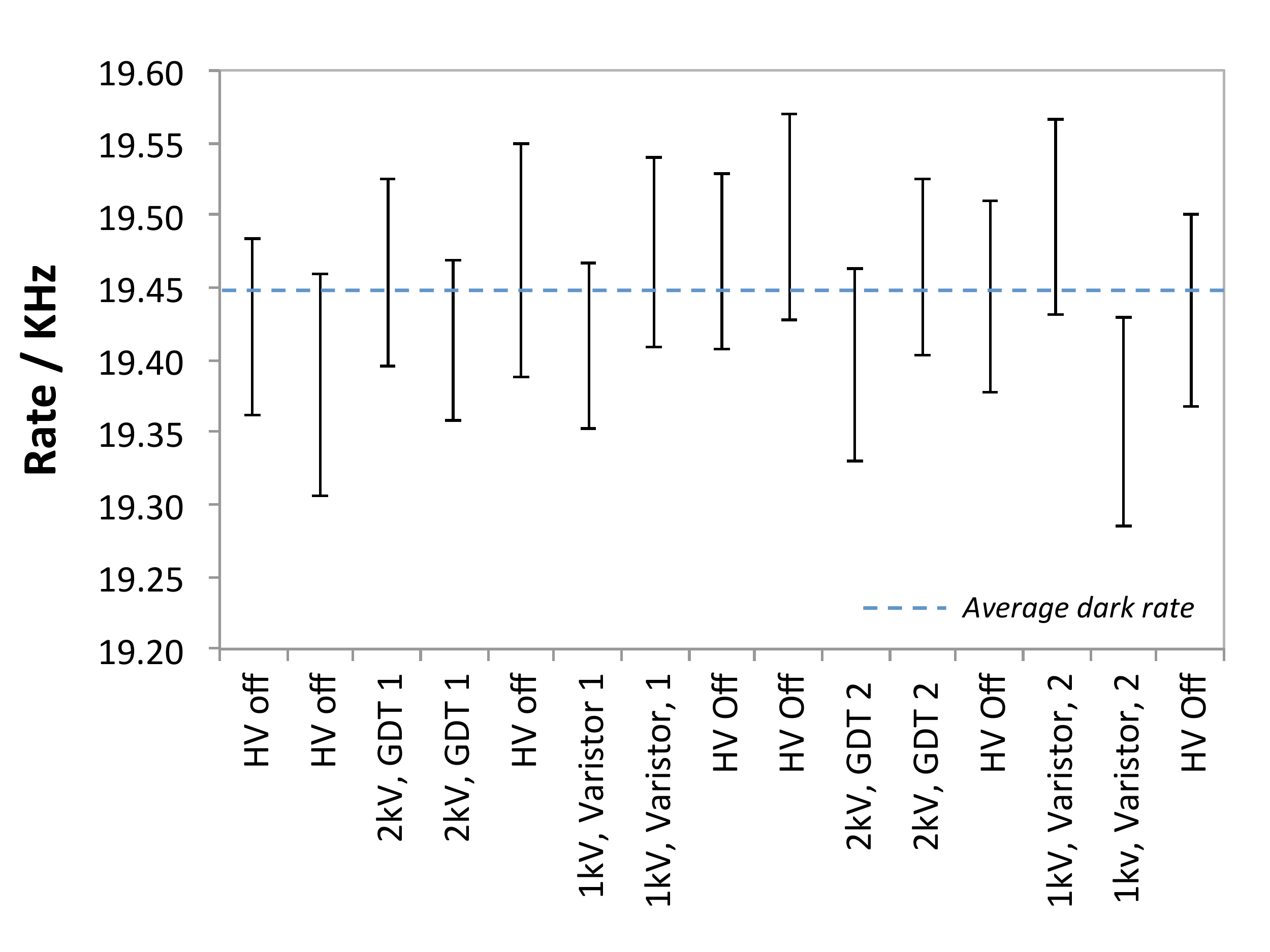}
\caption{The rate of a single photoelectron trigger with different devices held at nominal high voltage around 15~cm from an LBNE lightguide detector \label{fig:SPEPlot}}
\par\end{centering}
\end{figure}

The prototype optical detectors consist of an acrylic light guide coated in a wavelength shifting TPB coating ~\cite{Baptista:2012bf}.  This is coupled to three silicon photomultiplier (SiPM) detectors which have their signals fed to a shaper.  For this test, the shaped signal from one SiPM was viewed using an oscilloscope.  Binning the pulse amplitudes shows a clear separation of each $N$ photoelectron peaks up to at least 20.  The detectors have an estimated average collection efficiency of O(0.03\%) \cite{DenverPrivate} for 128~nm liquid argon scintillation light.  A trigger was set at 10~mV, which is below the single photoelectron level of 12~mV, and the single photoelectron trigger rate was determined by counting the number of triggers in five minutes.  The rate when no voltage was supplied to any surge protector was 19.44$\pm$0.02~kHz, which includes both thermionic SiPM noise and the prompt and late scintillation signals from cosmic ray muons passing through the detector.  The rate was compared to that measured when each device was biased at 2~kV for GDTs, or 1~kV for varistors.  No significant additional single photoelectron rate was observed, which sets an upper limit on the detected rate of 100~Hz at a distance of 15 cm, which is below the level where there would be interference with a LArTPC optical system.  These data are shown in Figure~\ref{fig:SPEPlot}.

\begin{figure}
\begin{centering}
\includegraphics[width=0.48\columnwidth]{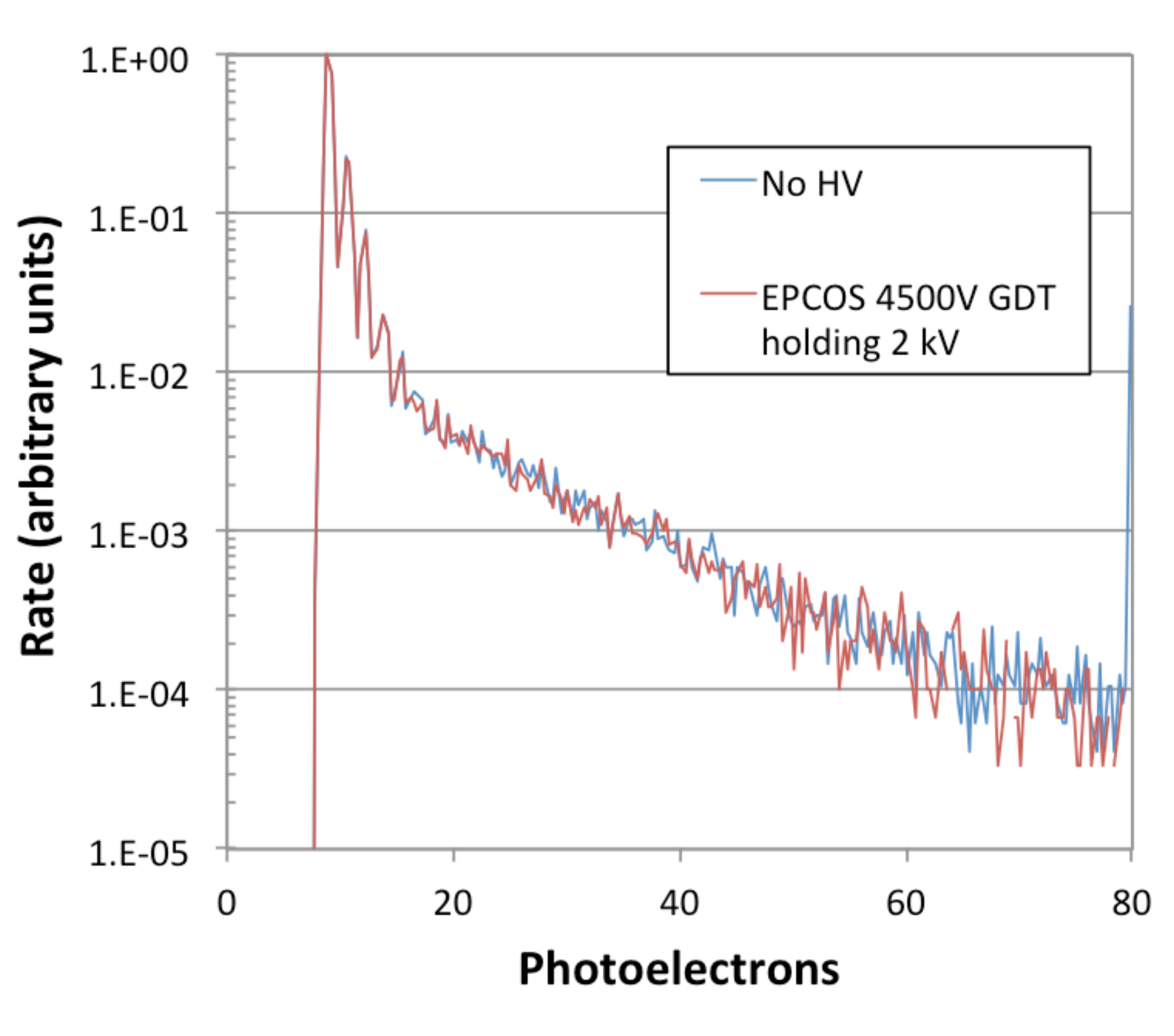}\includegraphics[width=0.48\columnwidth]{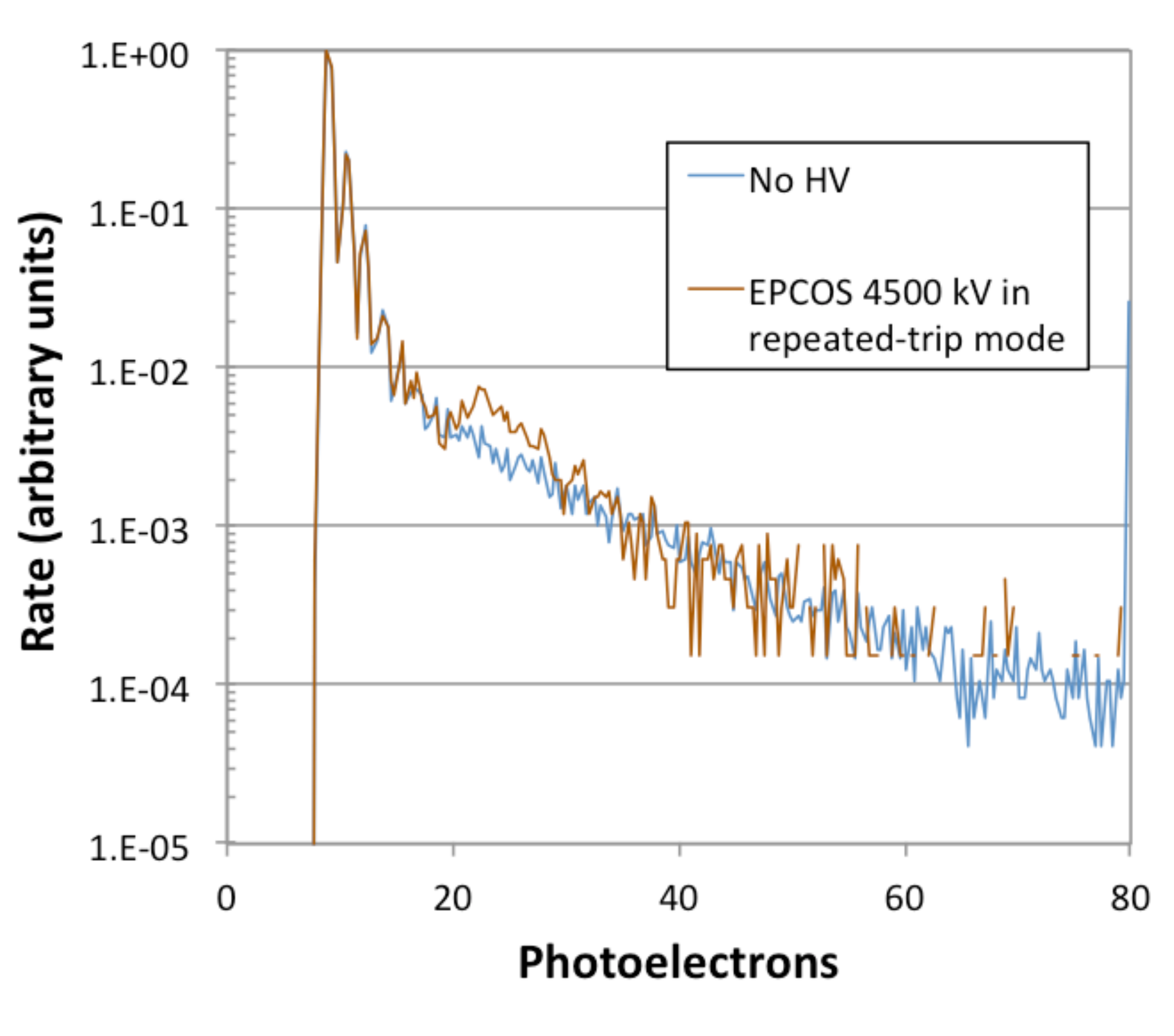}
\caption{The pulse amplitude spectra in both the nominal (left) and discharge (right) conditions, compared to the cosmic ray background in TallBo. \label{fig:LightSpectra}}
\par\end{centering}
\end{figure}

With the power supply set above the clamping point, light emission was observed for the GDTs, originating from the spark formed inside the device. No light emission was observed for varistors.  In this mode, the power supply repeatedly biased the device and then tripped, leading to short bursts of light.  The spectrum of pulses recorded from the SiPM for the GDT operated in nominal and over-voltage mode is shown in Figure~\ref{fig:LightSpectra}, left and right.  In the over-voltage mode, a distinct feature is seen at around 20 photoelectrons.  In a real transient over-voltage situation, this pulse is likely to be very small in comparison to the light emitted by the initial spark which caused the discharge.

\section{Conclusions} \label{sec:Conclusions}

We have demonstrated in this paper that high voltage GDTs and varistors continue to function as surge protection devices in a liquid argon environment.  In both cases, the devices become more insulating in the non-conducting regime, and continue to exhibit clamping behavior, albeit at a somewhat reduced clamping voltage.  For GDTs, the clamping voltage is seen to drop significantly to a value not strongly correlated with the warm clamping voltage, whereas for varistors the change is smaller, and there is a strong correlation between warm and cold clamping voltages.

The devices have also been demonstrated to be robust under application of the surge energies and voltages to be expected in the HV discharge of a current-generation LArTPC experiment.  Furthermore, they have been demonstrated to offer protection to a specific class of vulnerable components, the resistors originally selected to be used in the MicroBooNE field cage.  The possible effects of each device upon argon purity have been investigated and both are found to be benign.  Finally, both devices have been tested for light emission under nominal operating conditions and no large photon flux was observed.  Both classes of device have been shown to be appropriate for use in LArTPC experiments as surge protection devices.

When applied in parallel with a vulnerable component, both classes of surge protector prevent the voltage from rising above a critical clamping voltage for long timescales, ensuring that the sensitive component is not held in a large over-voltage condition.  Although both act as surge protectors, there are key differences in functionality between GDTs and varistors.  Varistors have a smooth I$-$V curve, free from hysteresis, producing a gradual clamping behavior.  An increasing current supply is met with a decreasing resistance, thus preventing a large potential difference from evolving.  GDTs, on the other hand, transition discretely from the open to conductive state when a critical voltage is reached, acting as a crowbar.  Current is allowed to flow through the GDT until the potential difference between the electrodes falls below the GDTs extinction voltage, which is significantly lower than the initial spark-over voltage.  This will occur when the current source driving the discharge is exhausted, which is expected to take on the order of microseconds in a TPC discharge event, assuming a suitable current limiting resistor is used in line with the HV power supply.

\begin{figure}[t]
\centering \includegraphics[width=0.98\textwidth]{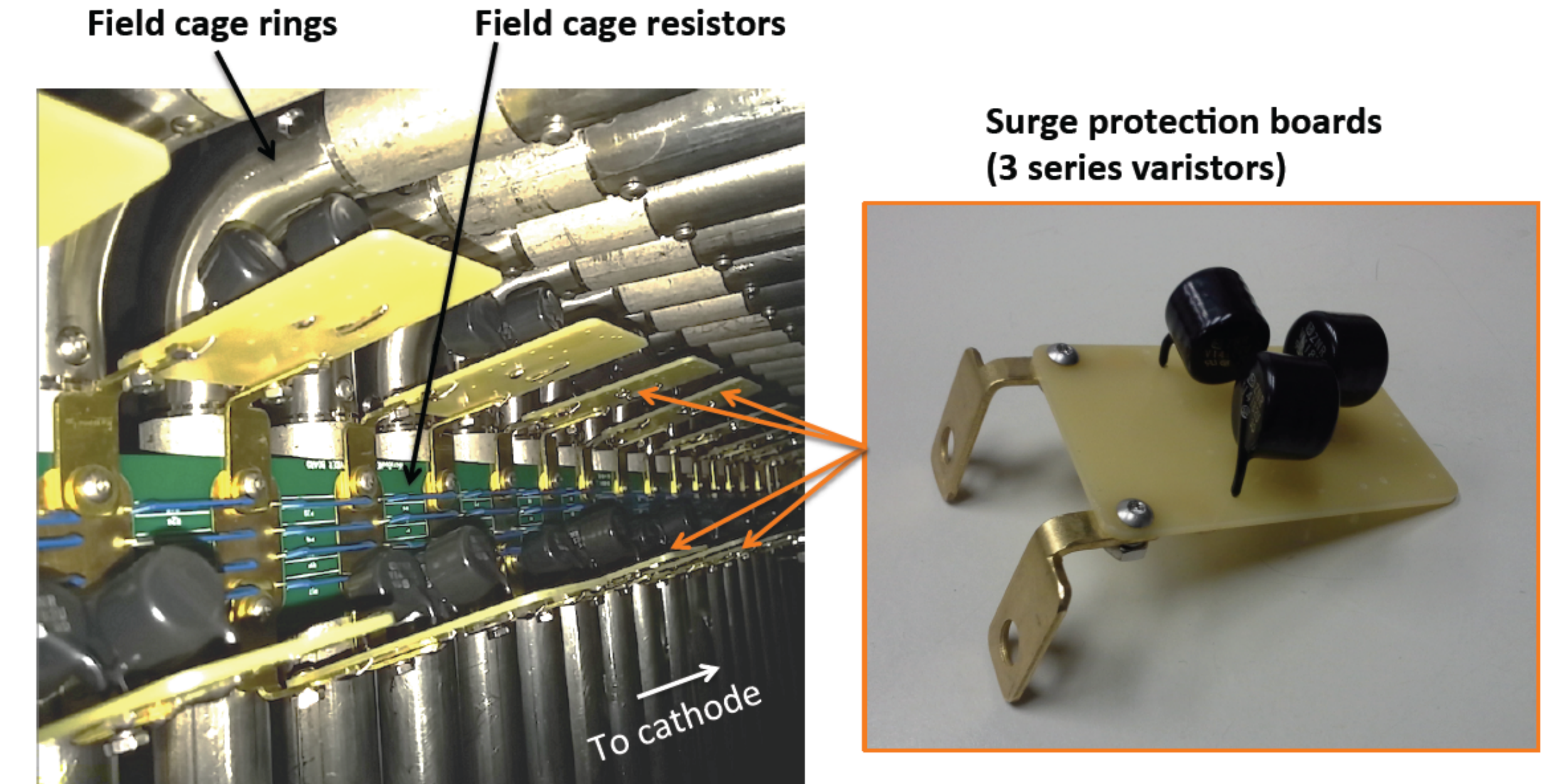}
\caption{The MicroBooNE surge protection system, as installed in the TPC.  Each set of field cage resistors between the cathode and field cage ring number 32 is placed in parallel with three series varistors.} 
\label{fig:AsInstalled} 
\end{figure}

Even with surge protectors applied, very short transients, with timescales on the order of the light travel time across the detector may be present.  Understanding these transients requires a detailed analysis of the inductive, capacitive and resistive couplings within the field cage.  The dynamics of these transients within the MicroBooNE TPC have been studied, but these studies are beyond the scope of this paper, and will be presented in a future publication~\cite{uBooNE_NIM}.  The role of a surge protector is to prevent such over-voltages from being maintained over timescales long enough to cause component damage, and in cases where the current supply driving the discharge is finite, as it is in a LArTPC spark condition, they perform this role in a relatively application-independent way.

There are advantages and disadvantages to both GDTs and varistors for high voltage surge protection.  In particular, easily mountable varistors are not commercially available in the appropriate voltage range for typical LArTPC applications, so several devices must be used in series.  This requires more solder connections and mounting hardware than in the case of a GDT, where a single unit may be mounted directly in parallel with the vulnerable component to be protected.  On the other hand, the gradual clamping behavior and well defined I$-$V curve of varistors makes it easier to model the dynamics of the field cage in surge conditions using software such as {\sc spice}~\cite{spice}.  

As a result of this work, a varistor based surge protection system was installed in parallel with the MicroBooNE TPC resistor chain between the the cathode and the thirty-two nearest field cage rings.  Sets of three series varistors, having a combined clamping voltage of $5.26 \pm 0.04$~kV, were mounted on G10 boards and affixed using brass contacts to the mount points of the field cage resistor boards on the TPC frame.  A picture is shown in Figure~\ref{fig:AsInstalled}.  

As well as being suitable for this specific application, this work demonstrates generally that GDTs and varistors continue to act as surge protection devices in liquid argon environments.  As such, both classes of device may find wider applications in electrical protection of sensitive components in LArTPC detectors.

\section*{Acknowledgements}

We thank Walter Jaskierny for assistance in the design and implementation of safe and conclusive high voltage tests, and Jonathan Woodworth of ArrestorWorks for helpful consultations on industrial use of varistors.  We thank Henning Back helping us to perform gas analysis using a UGA which is part of the low-radioactivity argon purification project, Stuart Mufson and Denver Whittington for the equipment and expertise used to measure light emission from devices in the Bo test stand, Stephen Pordes for allowing us access to the MTS and other facilities at the Proton Assembly Building, Ewa Skup for operating the MTS, and Sarah Lockwitz for setting up the MicroBooNE HV feedthrough test stand.  We also thank our MicroBooNE collaborators for giving helpful feedback at all stages of this work, in particular Linda Bagby, Regina Rameika, Jennifer Raaf, Anne Schukraft, and Michele Weber.

This work was supported by the Fermi National Accelerator Laboratory, which is operated by the Fermi Research Alliance, LLC under Contract No. De-AC02- 07CH11359 with the United States Department of Energy.  The surge protection components under test, and the work by BJPJ and JMC, were funded by the National Science Foundation grant PHY-1205175.  JA is supported by National Science Foundation grant PHY-1068553.  SG is supported by the Department of Energy through grant DE-FG03-99ER41093 and JMSJ through grant DE-SC0011784. TS acknowledges the support of the Swiss National Science Foundation.  JZ is supported by the University of Chicago.

\bibliographystyle{ieeetr}
\bibliography{SurgeProtection}
 
\end{document}